\documentclass[a4paper,traditabstract]{aa}
\usepackage[english]{babel}
\usepackage{amssymb}
\usepackage{subfig}
\usepackage[usenames,dvipsnames,svgnames,table]{xcolor}
\usepackage[bookmarks, colorlinks, breaklinks]{hyperref}  
\hypersetup{linkcolor=blue,citecolor=SteelBlue,filecolor=black,urlcolor=SteelBlue}

\usepackage[varg]{txfonts}
\usepackage{microtype}

\usepackage{latexsym}
\usepackage{natbib}
\bibpunct{(}{)}{;}{a}{}{,} 

\usepackage{siunitx}
\DeclareSIUnit\parsec{pc}
\DeclareSIUnit\erg{erg}
\DeclareSIUnit\deg{deg^{-2}}
\DeclareSIUnit\count{cnt}
\DeclareSIUnit\solarmass{M_{\odot}}

\def\xmm{XMM-{\it Newton}}
\def\chandra{{\it Chandra}}
\def\planck{{\it Planck}}
\def\nh{N_{\rm H}}

\def\MY {$M_{500}$--$Y_{\textrm X}$}

\def\YX {$Y_{\textrm X}$}
\def\TX {$T_{\rm X}$}
\def\Rv {$R_{\rm 500}$}

\def\blline{median}

\newcommand{\eqiac}[1]{Eq. \ref{#1}}
\newcommand{\seciac}[1]{Sect. \ref{#1}}
\newcommand{\tabiac}[1]{Table \ref{#1}}
\newcommand{\figiac}[1]{Fig. \ref{#1}}

\newcommand{\citeiac}[1]{(\citealt{#1})}

\newcommand{\inv}[1]{\phantom{#1}}

\newfont{\gwpfont}{cmssq8 scaled 1000}
\newcommand{\rexcess}{{\gwpfont REXCESS}}
\newcommand{\cccp}{{\gwpfont CCCP}}
\begin{document}

\title{Recovering galaxy cluster gas density profiles \\ with \xmm\ and \chandra}

\author{I. Bartalucci\inst{1,2} \and M. Arnaud\inst{1,2} \and G.W. Pratt\inst{1,2} \and A. Vikhlinin\inst{3}
 \and E. Pointecouteau\inst{4} \and W.R. Forman\inst{3} \and C. Jones\inst{3}
  \and P. Mazzotta\inst{3,5} \and F. Andrade-Santos\inst{3}} 

\institute{IRFU, CEA, Universit\'e Paris-Saclay, F-91191 Gif Sur Yvette, France
		  \and Universit\'e Paris Diderot, AIM, Sorbonne Paris Cit\'e, CEA, CNRS, F-91191 Gif-sur-Yvette, France
           \and Harvard-Smithsonian Center for Astrophysics, 60 Garden Street, Cambridge, MA 02138, USA 
           \and IRAP, Université de Toulouse, CNRS, UPS, CNES, Toulouse, France 
           \and Dipartimento di Fisica, Università di Roma Tor Vergata, via della Ricerca Scientifica 1, 00133 Roma, Italy}

\date{}

\abstract{
We examine the reconstruction of galaxy cluster radial density profiles obtained from \chandra\ and \xmm\ X-ray observations, using high quality data for a sample of twelve objects covering a range of morphologies and redshifts.
By comparing the results  obtained from the two observatories and by varying key aspects of the analysis procedure, we examine the impact of instrumental effects and of differences in the methodology used in the recovery of the density profiles.  We find that the final density profile shape is particularly robust. We adapt the photon weighting vignetting correction method developed for \xmm\ for use with \chandra\ data, and confirm that the resulting \chandra\ profiles are consistent with those corrected a posteriori for vignetting effects.
Profiles obtained from direct deprojection and those derived using parametric models are consistent at the $1\%$ level.  At radii larger than $\sim 6\arcsec$, the agreement between \chandra\ and \xmm\ is better than $1\%$, confirming an excellent understanding of the \xmm\ PSF. Furthermore, we find no significant energy dependence. 
The impact of the well-known offset between \chandra\ and \xmm\ gas temperature determinations on the density profiles is found to be negligible. However, we find an overall normalisation offset in density profiles of the order of $\sim 2.5\%$, which is linked to absolute flux cross-calibration issues. As a final result, the weighted ratios of \chandra\ to \xmm\ gas masses computed at $R_{2500}$ and $R_{500}$ are $r=1.03 \pm 0.01$ and $r=1.03 \pm 0.03$, respectively. Our study confirms that the radial density profiles are robustly recovered, and that any differences between \chandra\ and \xmm\ can be constrained to the $\sim 2.5\%$ level, regardless of the exact data analysis details. These encouraging results open the way for the true combination of X-ray observations of galaxy clusters, fully leveraging the high resolution of \chandra\ and the high throughput of \xmm. 
}

 \titlerunning{Recovering galaxy cluster gas density profiles with \xmm\ and \chandra}
\maketitle

\section{Introduction}\label{introduction}
Clusters of galaxies provide valuable information on cosmology, from the physics driving galaxy and structure formation to the nature of dark energy  \citep[see e.g.][]{voit2005,allen2011}. Clusters are primarily composed of dark matter, the baryonic component being contained mainly in the form of hot ionized plasma that fills the intra-cluster volume, namely the intra-cluster medium (ICM). The ICM emits in the X-ray band primarily via thermal Bremsstrahlung, which depends on the plasma density and temperature. 

X-ray observations play a key role in studying the ICM emission. 
In particular, the advent of high resolution spectro-imaging observations from \chandra\ and \xmm\ has yielded a rich ensemble of observations to be  studied and understood. In the context of galaxy cluster observations these two satellites are complementary: the unprecedented angular resolution of \chandra\ is well-suited to explore the bright central parts, while the large effective area of \xmm\ allows detection of the emission up to large radial distances, where the X-ray signal becomes very faint. A combination of these instruments can in principle be used to obtain more precise (especially in the centre) and spatially extended profiles \citep[e.g.][]{bar17}. However, previous works combining datasets from the two satellites have found systematic differences due to cross-calibration issues \citep[e.g.,][]{snowden2008, mahdavi2013, martino2014,sch2015}. For the hottest clusters, \chandra\ temperatures are found to be higher than those of \xmm\ by $10-15\%$ while the flux offset is smaller, at $\sim 1-3\%$, and consistent with zero. Furthermore, as instruments are routinely monitored throughout the spacecraft lifetime and as in-flight calibration is undertaken, these systematic differences can evolve considerably depending on the version of the calibration database, {\tt CALDB}, used for each satellite.

\begin{table*}[!ht]

\caption{{\footnotesize Observational properties of the twelve clusters used in this work. The coordinates indicate the position of the X-ray peak used as the centre for profile extraction. The peak was determined using \chandra\ observations in the $[0.7-2.5]$ keV band.}}\label{tab:sample}

\begin{center}

\resizebox{\textwidth}{!} {

\begin{tabular}{lrrccccccc}
\hline        
\hline

\multicolumn{1}{c}{Cluster name} & \multicolumn{1}{c}{RA} &

\multicolumn{1}{c}{DEC} & \multicolumn{1}{c}{$z$} &

\multicolumn{1}{c}{$\nh\,^a$} & \multicolumn{1}{c}{\chandra\ } &

\multicolumn{1}{c}{\xmm\ } &

\multicolumn{1}{c}{\chandra} & \multicolumn{1}{c}{\xmm} \\

\multicolumn{1}{c}{} & \multicolumn{1}{c}{} &

\multicolumn{1}{c}{} & \multicolumn{1}{c}{} &

\multicolumn{1}{c}{} & \multicolumn{1}{c}{exp. } &

\multicolumn{1}{c}{exp. (MOS1,2; PN)} &

\multicolumn{1}{c}{obs. Id 	} & \multicolumn{1}{c}{obs. Id 	} \\

\multicolumn{1}{c}{} & \multicolumn{1}{c}{$[J2000]$} &

\multicolumn{1}{c}{$[J2000]$} & \multicolumn{1}{c}{} &

\multicolumn{1}{c}{$[10^{20}\si{\centi\metre}^{-3}]$} & \multicolumn{1}{c}{[\si{\kilo\second}] } &

\multicolumn{1}{c}{[\si{\kilo\second}]} &

\multicolumn{1}{c}{} & \multicolumn{1}{c}{} \\

\hline

A1651               					& $12$ $59$ $22.15$ 		& $-04$ $11$ $48.24$  	 	& $0.084$    &  $1.81$                         				&$\inv{0}10$          		& $\inv{00}7$; $\inv{00}5\inv{0}$ 			& $\inv{0}4185$		&$0203020101$			\\

A1650               					& $12$ $58$ $41.48$ 		& $-01$ $45$ $40.45$ 		& $0.084$    	&  $0.72$                         				&$\inv{0}37$          		& $\inv{0}34$; $\inv{0}30\inv{0}$ 			& $\inv{0}7242$		&$0093200101$			\\

A1413               					& $11$ $55$ $17.97$     	& $+23$ $24$ $21.99$	 	& $0.143$    	&  $1.84^{*}$                         		&$\inv{0}75$      			& $\inv{0}63$; $\inv{0}52\inv{0}$ 			& $\inv{0}5003$		&$0502690201$			\\

A2204             					& $16$ $32$ $46.94$  	& $+05$ $34$ $32.63$  	& $0.152$    	&  $6.97^{*}$                         		&$\inv{0}77$     			& $\inv{0}14$; $\inv{0}10\inv{0}$ 			& $\inv{0}7940$		&$0306490201$			\\

A2163   	           				& $16$ $15$ $46.06$ 		& $-06$ $08$ $52.36$		& $0.203$    	&  $16.50^{*}$                           	&$\inv{0}71$          		& $\inv{0}92$; $\inv{0}65^{\dagger}$ & $\inv{0}1653$		&$0694500101$			\\ %

A2390              					& $21$ $53$ $36.82$ 		& $+17$ $41$ $43.30$	 	& $0.231$    &  $8.66$                         				&$\inv{0}95$          		& $\inv{0}10$; $\inv{00}8\inv{0}$ 			& $\inv{0}4193^{\bullet}$		&$0111270101$			\\

MACS J1423.8+2404  		& $14$ $23$ $47.92$	 	& $+24$ $04$ $42.96$	 	& $0.545$    &  $2.20$                         				&$115$         			    & $\inv{0}34$; $\inv{0}22^{\dagger}$  & $\inv{0}4195^{\bullet}$		&$0720700301$	\\  %mos2 43

MACS J0717.5+3745$^b$	& $07$ $17$ $31.75$	 	& $+37$ $45$ $31.02$	 	& $0.546$    &  $6.64$                         				&$\inv{0}58$          		& $\inv{0}59$; $\inv{0}52\inv{0}$ 			& $16305$ 				&$0672420201$			\\

MACS J0744.9+3927			& $07$ $44$ $52.78$		& $+39$ $27$ $26.63$		& $0.698$	    & $5.66$									&$\inv{0}50$ 				&$\inv{0}52$; $\inv{0}42^{\dagger}$ 	&$\inv{0}6111$ 		&$0551851201$			\\ %

SPT-CL J2146-4633			& $21$ $46$ $34.72$	 	& $-46$ $32$ $50.86$	& $0.933$    &  $1.64$                         				&$\inv{0}80$          		& $\inv{0}90$; $\inv{0}64^{\dagger}$ & $13469$			&$0744400501$	\\ %

SPT-CL J2341-5119			& $23$ $41$ $12.23$	 	& $-51$ $19$ $43.05$	& $1.003$    &  $1.21$                         				&$\inv{0}50$          		& $\inv{0}64$; $\inv{0}34^{\dagger}$ 	& $11799$				&$0744400401$	\\    %63 M1, 67 M2

SPT-CL J0546-5345			& $05$ $46$ $37.22$	 	& $-53$ $45$ $34.43$	& $1.066$    &  $6.77$                         				&$\inv{0}27$          		& $\inv{0}45$; $\inv{0}33^{\dagger}$ 	& $\inv{0}9336$		&$0744400201$	\\ %

\hline

\end{tabular}

}

\end{center}

\footnotesize{Notes: $^a$ The neutral hydrogen column density was derived from the LAB survey \citep{kalberla2005}. $^b$ The X-ray peak is centred on the main structure, identified as region "B" in \citealt{limousin2016}.  $^{*}$the absorption value was estimated in a region which maximises the signal to noise. $^{\circ}$ MOS 2 effective exposure times are $3-4 \%$ higher than MOS1, except for MACS\, J1423.8+2404, for which the MOS2 exposure time is $43$ ks. $^\bullet$ ACIS-S observations.}

\end{table*}

%---

Here we focus uniquely on the reconstruction of gas density profiles and the calculation of integrated gas masses from \xmm\ and \chandra\ data. The gas density should, in principle, be trivial to recover since the observed X-ray surface brightness is proportional to the line-of-sight integral of the square of the density,  times the X-ray cooling function, which in turn is nearly independent of temperature in the soft energy band. 
Thus the absolute temperature calibration will have a minimal effect on the recovered density (we show explicitly that this is the case in Sect.~\ref{lambda_study}). However, a number of other potential sources of systematic differences still remain, including the absolute flux calibration, the exclusion of substructure and point sources, the correction for the point spread function (for \xmm), the method used to correct for the vignetting of the X-ray telescopes, or whether the analysis uses a parametric or a non-parametric approach. 
Here we compare the ICM density profiles obtained using variations on two approaches developed for statistical studies of clusters -- that of the Chandra Cluster Cosmology Project \citep[\cccp; ][]{vikh2009} for \chandra, and that of \rexcess\ \citep[][]{croston2008,pratt2009} for \xmm. The Planck Collaboration also used the latter method \citep[e.g.][]{PERxmmesz,PIPmass,P13szcount}. Our choice is motivated by the fact that each approach is representative of a standard analysis for its respective satellite, while being sufficiently different in procedure (notably in the treatment of vignetting, or in the choice of parametric vs non-parametric modelling of the X-ray surface brightness) to allow investigation of some of the issues mentioned above. We can also use the  exceptional PSF of \chandra\ to test the robustness of the \xmm\ PSF correction procedure, which is one of the main analysis issues when reconstructing the central regions of the gas density profiles. This is particularly true for evolution studies, because of the decrease of the angular size with the redshift. Hence we focus on a sample in a wide redshift range, up to $z\sim1$.

The paper is organised as follows: Sect.~\ref{Data} presents the sample of the clusters used in this work, and Sect.~\ref{sec_analysis} 
details the various data analysis steps and methods used to reconstruct the radial density distribution  from the surface brightness profiles. In Sect.~\ref{robustness}, we describe the tests we use to asses and quantify the robustness of the density profile reconstruction, and detail the comparison between the two instruments. Our conclusions are presented in Sect.~\ref{conclusions}.

We adopt a cold dark matter cosmology with $\Omega_{M}=0.3$, $\Omega_{\Lambda}=0.7$ and $H_{0} = \SI{70}{\kilo\meter \per\mega\parsec \per\second }$ throughout. All errors are reported at the $1\sigma$ level. The quantity \Rv\ is defined as the radius at which the total density of the cluster is $500$ times the critical density.

%%%%%%%%%%%%%%%%%%%%%%%%%%%%%%%%%%%%%%%%%%%%%%%%%%%%%%%%%%%%%%%%%%%%%%%%%%
\section{Cluster sample}\label{Data}

As this work is dedicated to a systematic comparison of the ICM properties derived from \chandra\ and \xmm\ observations, we require (1) sufficient exposure time to allow extraction of well-sampled radial profiles at least up to $\sim R_{500}$; (2) a wide redshift range to test the effect of different angular sizes on the PSF reconstruction; (3) different X-ray morphologies to examine the effect of features such as e.g. peaked or flat central emission. 
Following these criteria we define a sample of twelve clusters whose main observational properties are reported in Table~\ref{tab:sample}. We limit our datasets to single observations to simplify the analysis and avoid complications related to the creation of mosaics.

All the clusters in our sample have been observed using the  \chandra\ Advanced CCD Imaging Spectrometer (ACIS, \citealt{garmire2003}) and the \xmm\ European Photon Imaging Camera (EPIC). \chandra\ operates using a combination of two CCD arrays where the focus can be placed on either, the ACIS-Imaging or the ACIS-Spectroscopy (ACIS-I and ACIS-S observations, respectively). The latter instrument is composed of three cameras, MOS$1,2$ \citep{turner2001} and PN \citep{struder2001}, which operate simultaneously. 

\chandra\ observations of A2390 and MACSJ1423.8+2404 are ACIS-S, all the others are ACIS-I.

The \chandra\ observations of the six nearest clusters in Table~\ref{tab:sample} have previously been analysed in \citet[][hereafter V06 and V09]{vikhlinin2006,vikh2009} using the \cccp\ procedure. The corresponding \xmm\ observations were analyzed in \citet{PERxmmesz} using the \rexcess\ analysis procedures. 
These published results are directly compared in Sect.~\ref{first_comp}.

%%%%%%%%%%%%%%%%%%%%%%%%%%%%%%%%%%%%%%%%%%%%%%%%%%%%%%%%%%%%%%%%%%%%%%%%%%
\section{Data analysis}\label{sec_analysis}

%%%%%%%%%%%%%%%%%%%%%%%%%%%%%%%%%%%%%%%%%%%%%%%%%%%%%%%%%%%%%%%%%%%%%%%%%%
Appendix~\ref{appendix_reduction} summarises the basic data reduction, including production of calibrated event files, point source masking and background estimation.

\subsection{Effective area  correction }\label{sec:weight}
To derive the density profile, we need to convert count rates to  emission measure ${\rm EM}=\int n_{\rm e}^2~dv$. These quantities are related by the cooling function in the energy band under consideration, $\Lambda(T,Z)$, taking into account  the absorption, the redshift, and the instrumental  response (i.e. the effective area as a function of energy and position). $\Lambda(T,Z)$ depends on the temperature and metallicity in the region under analysis. The effective areas of \chandra\ and \xmm\ vary across the field of view as a result of different effects, including the non-uniformity of the CCD quantum efficiency, and the telescope vignetting (which reduces the number of photons detected on different portions of the detector). These spatial effects depend on the photon energy, a measure that itself is affected by uncertainty. 

As discussed in \citet{arnaud2001}, there are two procedures that are generally used to take into account vignetting effects. The first, which we call the post-correction method, and which was adopted by V06/V09 and in the \cccp\ analysis, consists of computing the effective area at each position. The area-averaged effective area for each annulus is used to convert the count rate  to emission measure along the line of sight. 

The second, which we call the pre-correction method \citep[][]{arnaud2001,churazov2008}, and which was adopted in the \rexcess\ analysis, first corrects the data for the spatial dependence of the instrument response. 
For each photon, detected at the position $(i,j)$ and with energy $e$, a weight $W(i,j,e)$ is assigned such that:
\begin{equation}\label{eq:wght_xmm}
W(i,j,e) = \frac{A_{\rm eff}\,(i_0,j_0,e)}{A_{\rm eff}\,(i,j,e)},
\end{equation}
where $A_{\rm eff}\,(i,j,e)$ is the effective area computed at the photon position and $A_{\rm eff}\,(i_0,j_0,e)$ is the effective area at the aim-point, i.e. where it attains its maximum value. 
The corrected number counts in a given extraction region and energy band is then simply the sum of the weights. This represents the number of photons we would obtain if the detector had a uniform response equivalent to that at the aim-point position. 

The pre-correction method was proposed by  \citet{arnaud2001}  primarily for spectroscopic analysis in cases  when the spatial distribution of the source cannot be considered as uniform across the extraction region. In such cases, the averaging of the effective area over the region under consideration is not trivial and is potentially  biased. For surface brightness profile studies, the main advantage of the method is the simplification of the analysis (particularly for the cosmic background subtraction). The main drawback is a statistical degradation as compared to the direct method, by a typical 
 factor of $\sqrt{\langle{W^2}\rangle}/\langle{W}\rangle$, where  the brackets denote the average over the extracted photons. However, this degradation is limited as the vignetting variations are small across typical annulus widths and in the soft energy band. Note also that the method is only exact for a perfect knowledge of the photon energy and position. In practice, it requires that the vignetting does not vary significantly within the PSF scale and within the uncertainties on the photon energy (i.e. the spectral resolution). This is the case both for \chandra\ and \xmm. 

For \xmm\ we computed the weights using the built-in SAS \verb?evigweight? routine. For \chandra, which has no built-in function, we have developed a procedure that computes an analogous weight quantity using CIAO tools.
The procedure is described in detail in 
Appendix \ref{appendix_weight}. In addition to the weight, for each photon we also compute the exposure time using the related exposure maps generated for \chandra\ and \xmm\ using the \verb?fluximage? and the \verb?eexpmap? tools, respectively.

To determine the impact of these different vignetting correction techniques, in \seciac{robustness} we compare the density profiles obtained from the pre-correction method and from the reference V06/V09  profiles calculated with the post-correction procedure. 

\bgroup
\def\arraystretch{1.2}%  1 is the default, change whatever you need
\begin{table*}[t]
\caption{{\footnotesize Table of gas masses computed at fixed radii. Here $R_{500,Yx}$ is measured via the \MY\ relation, using the \xmm\ datasets.}}\label{tab:500_prop}
\begin{center}
\resizebox{2.\columnwidth}{!} {
\begin{tabular}{lccccccccc}
\hline        
\hline
\multicolumn{1}{c}{Cluster name} & \multicolumn{1}{c}{$R_{500,Yx}$}  & \multicolumn{4}{c}{$M_{gas} \; (R<R_{2500})$} & \multicolumn{4}{c}{ $M_{gas} \; (R<R_{500})$} \\

                    				&              					&        XMM       		 			&     		CXO					  &   XMM PXI         	  & CXO V06/V09 		   & XMM 						& CXO  						&      XMM PXI           &  CXO V06/V09		\\
\multicolumn{1}{c}{ } & \multicolumn{1}{c}{ [kpc] } & \multicolumn{4}{c}{[$10^{13} M_{\odot}$]} & \multicolumn{4}{c}{[$10^{13} M_{\odot}$]}\\       				
\hline
A1651  &$1132_{-9}^{+9}$ &$ 2.01 \pm  0.01 $ &$ 2.10 \pm  0.01 $ &$ 2.10 \pm  0.04 $ &$ 2.03 \pm  0.01 $ &$ 5.61 \pm  0.05 $ &$ 6.05 \pm  0.07 $ &$ 6.11 \pm  0.07$ &$ 5.61 \pm  0.08 $ \\
A1650  &$1114_{-4}^{+4}$ &$ 1.81 \pm  0.00 $ &$ 1.87 \pm  0.01 $ &$ 1.83 \pm  0.02 $ &$ 1.84 \pm  0.00 $ &$ 5.16 \pm  0.02 $ &$ 5.11 \pm  0.06 $ &$ 5.26 \pm  0.06$ &$ 5.13 \pm  0.03 $ \\
A1413  &$1233_{-4}^{+4}$ &$ 2.86 \pm  0.01 $ &$ 2.97 \pm  0.01 $ &$ 2.90 \pm  0.03 $ &$ 2.88 \pm  0.01 $ &$ 7.87 \pm  0.03 $ &$ 7.89 \pm  0.05 $ &$ 8.02 \pm  0.05$ &$ 8.09 \pm  0.02 $ \\
A2204  &$1348_{-12}^{+12}$ &$ 4.01 \pm  0.02 $ &$ 4.22 \pm  0.01 $ &$ 4.07 \pm  0.03 $ &$ 4.04 \pm  0.02 $ &$10.94 \pm  0.12 $ &$11.54 \pm  0.08 $ &$11.49 \pm  0.08$ &$10.96 \pm  0.13 $ \\
A2163  &$1787_{-6}^{+6}$ &$10.17 \pm  0.02 $ &$10.42 \pm  0.03 $ &$10.41 \pm  0.09 $ &$ 9.92 \pm  0.04 $ &$32.63 \pm  0.09 $ &$34.12 \pm  0.14 $ &$31.26 \pm  0.13$ &$32.14 \pm  0.27 $ \\
A2390  &$1441_{-12}^{+14}$ &$ 5.85 \pm  0.03 $ &$ 6.04 \pm  0.02 $ &$ 5.57 \pm  0.05 $ &$ 5.88 \pm  0.03 $ &$15.50 \pm  0.12 $ &$16.54 \pm  0.09 $ &$15.86 \pm  0.09$ &$15.90 \pm  0.15 $ \\
MACS J1423.8+2404  &$981_{-13}^{+12}$ &$ 2.22 \pm  0.03 $ &$ 2.39 \pm  0.02 $ &- &-  &$ 6.31 \pm  0.09 $ &$ 6.74 \pm  0.08 $ &- &-  \\
MACS J0717.5+3745  &$1307_{-11}^{+12}$ &$ 5.72 \pm  0.03 $ &$ 5.83 \pm  0.05 $ &- &-  &$19.62 \pm  0.12 $ &$19.29 \pm  0.27 $ &- &-  \\
MACS J0744.9+3927  &$1032_{-11}^{+10}$ &$ 3.44 \pm  0.03 $ &$ 3.60 \pm  0.05 $ &- &-  &$10.69 \pm  0.11 $ &$10.70 \pm  0.19 $ &- &-  \\
SPT-CL J2146-4633  &$737_{-13}^{+12}$ &$ 1.00 \pm  0.01 $ &$ 1.02 \pm  0.03 $ &- &-  &$ 4.73 \pm  0.05 $ &$ 4.84 \pm  0.18 $ &- &-  \\
SPT-CL J2341-5119  &$769_{-14}^{+14}$ &$ 1.69 \pm  0.02 $ &$ 1.73 \pm  0.04 $ &- &-  &$ 5.12 \pm  0.08 $ &$ 5.19 \pm  0.18 $ &- &-  \\
SPT-CL J0546-5345  &$782_{-14}^{+17}$ &$ 1.71 \pm  0.02 $ &$ 1.82 \pm  0.06 $ &- &-  &$ 5.87 \pm  0.09 $ &$ 6.15 \pm  0.21 $ &- &-  \\

\hline
\end{tabular}
}
\end{center}
\end{table*}
\egroup
%%%%%%%%%%%%%%%%%%%%%%%%%%%%%%%%%%%%%%%%%%%%%%%%%%%%%%%%%%%%%%%%%%%%%%%%%%

\subsection{Surface brightness profiles}\label{surface_bright}

We extracted surface brightness profiles from concentric annuli centred on the X-ray peak in the $[0.3-2.5]\, \si{\kilo\electronvolt}$ and $[0.7-2]\, \si{\kilo\electronvolt}$ energy bands from \xmm\ and \chandra\ datasets, respectively. 
The minimum width of the annuli ($3.3\arcsec$ for \xmm; $2\arcsec$ for \chandra) was set in order to ensure a sufficient number of counts in each radial bin and to best exploit the resolution of the instruments. 
The count rate for each annulus was then computed as the sum of the weights divided by the appropriate exposure time, normalised by the area \citep{arnaud2001}. We removed the particle background by extracting the surface brightness profile of the recast CLOSED (\xmm) or STOWED (\chandra) event lists in the same annular regions, and subtracting these from the source surface brightness profile. 
We then identified a region free from cluster emission (generally an annulus in the outer regions of the observation) and determined the mean count-rate corresponding to the sky background level. This value was subtracted, and the profile re-binned to have a significance of $3\sigma$ per bin, using a logarithmic binning  factor of $r_{\rm out} / r_{\rm in} = 1.05$.

%%%%%%%%%%%%%%%%%%%%%%%%%%%%%%%%%%%%%%%%%%%%%%%%%%%%%%%%%%%%%%%%%%%%%%%%%%
\subsection{Temperature profiles}\label{emission_measure}

We undertook the following spectral analysis to estimate temperatures, which are needed to  compute the cooling function $\Lambda(T,Z)$. The particle background contribution was removed by subtracting the spectrum extracted from the background datasets in the same detector region.  
To model the sky background, we extracted the particle-background-subtracted spectrum from the region free of cluster emission defined in \seciac{surface_bright}.  We fitted this spectrum with a model consisting of an absorbed \citep[WABS;][]{morrison1983}
power law with index $\alpha=1.42$ \citep{lumb2002} for the extra-galactic Cosmic X-ray Background (CXB) emission, plus two absorbed {\sc mekal}  \citep{kaastra1992, liedhal1995} models for the Galactic emission \citeiac{kuntzsnowden2000}. The absorbing column density  was fixed to $N_{\rm H} = 0.7 \times 10^{20}$ cm$^{-3}$ \citep{parmar1999} for one of  the local Galactic component, and to the  21 cm value (\tabiac{tab:sample}) for the other two components.
The normalisations of all the components and the two {\sc mekal} temperatures were determined from the fit.
The sky background model was fixed to its best-fitting parameters and simply scaled to match the geometrical area of the region of interest. 
 We fitted an absorbed {\sc mekal} model, in addition to the fixed sky background model, to estimate the temperature of the ICM. The absorption was fixed to the Galactic line of sight value given \tabiac{tab:sample}. The abundance was left as a free parameter if the relative error was less than $30\%$, otherwise it was fixed to $0.3$.  The fits were performed in the $[0.7-10]\, \si{\kilo\electronvolt}$ and $[0.3-11]\, \si{\kilo\electronvolt}$ bands for \chandra\ and \xmm, respectively. For the \xmm\ fit, we also excluded the $ [1.4-1.6]\, \si{\kilo\electronvolt}$ band for all three cameras and the $[7.45-9.0]\, \si{\kilo\electronvolt}$ band for PN only \citep[see e.g.][]{pratt2007} to avoid prominent instrumental emission lines. 
All the models considered in the spectral analysis were convolved by the appropriate response functions. For \chandra\ the response matrix file (RMF) and the ancillary response file (ARF) were computed using the CIAO \verb?mkacisrmf? and \verb?mkarf? tools, respectively. The same was done for \xmm, using the SAS tools \verb?rmfgen? and \verb?arfgen?.
  
  %%%%%%%%%%%%%%%%%%%%%%%%%%%%%%%%%%%%%%%%%%%%%%%%%%%%%%%%%%%%%%%%%%%%%%%%%%

To determine the projected (2D) temperature profile, we extracted spectra from concentric annular regions centred on the X-ray peak, each annulus being defined to have $r_{\rm out} / r_{\rm in} = 1.3$ and a signal-to-noise ratio of at least $30$. After instrumental background subtraction, each spectrum was rebinned  to have at least $25$ counts per bin. We were able to measure a temperature profile for the nine lowest-redshift objects both for the \chandra\ and the \xmm\ samples.

\subsection{Density and gas mass profiles}\label{density_profiles}

We considered two different techniques to derive the radial (3D) density profiles from the X-ray surface brightness: parametric modelling, and non-parametric deconvolution and deprojection. The non-parametric deconvolution technique with regularisation is that proposed by \citet{croston2006}, which was used throughout the \rexcess\ analysis. Hereafter, we refer to this method as the deprojection technique, and refer to the derived density profiles from this analysis as the deprojected density profiles. Briefly, the observed surface brightness profile, $C_{\rm obs}$, can be modelled as the result of the emission produced by the hot plasma in concentric shells, $S_{\rm emit}$, projected along the line of sight, and convolved with the instrument PSF:
\begin{equation}\label{eq:croston}
[C_{\rm obs}] = [R_{\rm PSF}][R_{\rm proj}][S_{\rm emit}], 
\end{equation}
where $R_{\rm proj}$ and $R_{\rm PSF}$ are the projection and PSF matrices, respectively. The method consists of inverting \eqiac{eq:croston},
and using regularisation criteria to avoid noise amplification \citep[see][for details]{croston2006}. We used the analytical model described in \cite{ghizzardi2001} for the \xmm\ PSF. For our purposes the \chandra\ PSF can be neglected since the annuli used to perform imaging and spectroscopic analysis are far larger than the \chandra\ PSF over the entire radial range under consideration. We thus assumed an ideal PSF, i.e. the $R_{\rm PSF}$ for \chandra\ datasets is the unity matrix. 
The projection matrix $R_{\rm proj}$ was evaluated assuming that the emission comes from concentric spherical shells, each with constant density. The recovered density in the shell is thus the square root of the mean density squared. The contribution  from shells external to the maximum radius used for the surface brightness profile extraction was removed by modelling the deprojected surface brightness emission in the external regions with a power law. The deprojection matrix is purely geometrical and is the same for both \chandra\ and \xmm\ datasets.
Individual deprojected density profiles derived from the \chandra\ and \xmm\ datasets are shown in the upper panels of \figiac{fig:neall} with black solid lines and blue points, respectively.

We also derived the density profile using the parametric modified-$\beta$ model described in \citet{vikhlinin2006}, which we refer to hereafter as the parametric fit technique (the corresponding density profiles will be referred as parametric density profiles). Parameters were estimated through a fit of the observed surface brightness profile with a projected analytical model, convolved with the PSF. 
For clarity, in the figures, profiles have the suffix ${\rm dep}$ and ${\rm par}$, depending on whether the  deprojection or parametric technique has been used.

The errors were computed from $1000$ Monte Carlo realisations of the surface brightness. Unless otherwise stated, the cooling function profile used to convert to density was determined from the $2D$ temperature profile. For the three most distant clusters, the observation depth does not allow the determination of the temperature profile,  so we used a constant temperature measured in the region that maximises the signal to noise. 

We obtained gas mass profiles by integrating the deprojected density profiles in spherical shells. These are shown in \figiac{fig:mgasall} as black solid lines and blue points for the \chandra\ and \xmm\ datasets, respectively. We determined $R_{500}$ iteratively using the \MY\ relation of \citet{arnaud2010}, where \YX\ is the product of \TX, the temperature measured within $[0.15-0.75]R_{500}$, and $M_{\rm gas,500}$ \citep{krav2006}. Gas masses were calculated within this same aperture for both \xmm\ and \chandra\ datasets. The corresponding values are reported in \tabiac{tab:500_prop}.
While in principle we could use the hydrostatic mass profiles to obtain $R_{500}$ we choose instead the \YX\ proxy. This is expected to be more robust to the presence of dynamical disturbance  \citep[e.g.][]{krav2006} and allows us to obtain homogeneous mass estimates in the presence of lower quality data i.e. high-z objects.

%============================================================================================================================================================
\section{Robustness of density profile reconstruction}\label{robustness}

In the following we compare individual profiles by computing their ratio, after interpolation onto a common, regular grid in log-log space. We compute a `mean' ratio profile by taking either the median or the error weighed mean, and also calculate the $68\%$ confidence envelope. The two averaging methods give consistent results, and the results quoted below are defined with respect to the median profile. 

%%%%%%%%%%%%%%%%%%%%%%%%%%%%%%%%%%%%%%%%%%%%%%%%%%%%%%%%%%%%%%%%%%%%%%%%%%
\begin{figure}[t]
\begin{center}
\includegraphics[width=0.45\textwidth]{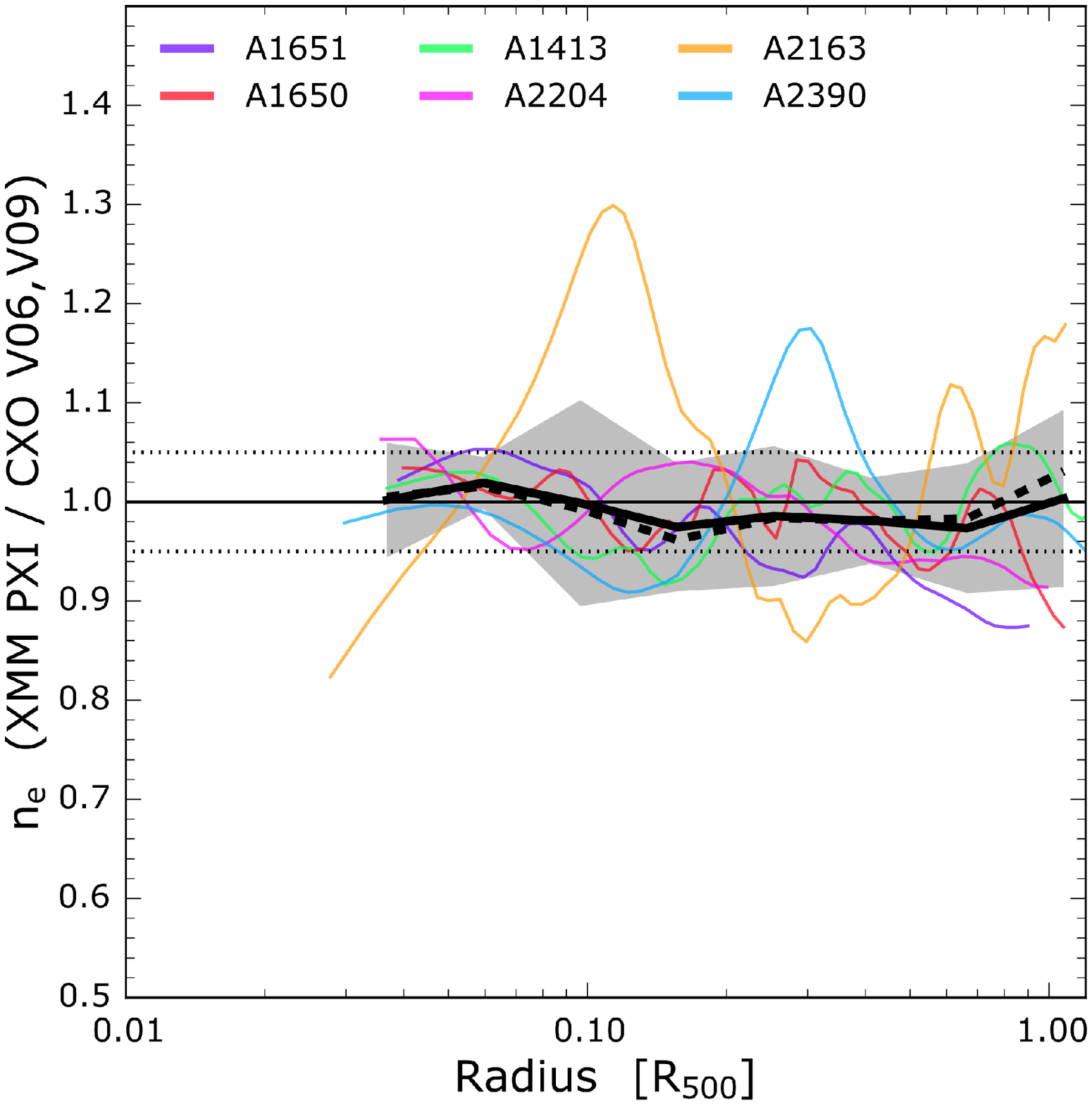} 
\caption{\footnotesize Ratio of \xmm\ and \chandra\ gas density profiles published by \citet{vikhlinin2006,vikh2009} and \citet{PERxmmesz} (PXI), obtained using independent analysis techniques. The \xmm\ and the \chandra\ profiles are derived from deconvolution/deprojection and parametric model fitting, respectively. The vignetting correction method is different (see text). The point source and substructure masking and centre choices are also independent. The black solid (dashed) line represents the \blline\ (weighed mean) and the grey shaded area its $1\sigma$ confidence level. The dotted horizontal lines represent the $\pm 5\%$ levels. 
At large scale the profiles show good agreement in the shape, with an average offset of $\sim2\%$. In the inner regions differences are dominated by the choice of the centre. The difference between the centres ranges from $1$ to $\sim 14$ arcsec ($<0.02 R_{500}$ in all cases).
 }
\label{fig:cons_dens}
\end{center}
\end{figure}
%%%%%%%%%%%%%%%%%%%%%%%%%%%%%%%%%%%%%%%%%%%%%%%%%%%%%%%%%%%%%%%%%%%%%%%%%%

%%%%%%%%%%%%%%%%%%%%%%%%%%%%%%%%%%%%%%%%%%%%%%%%%%%%%%%%%%%%%%%%%%%%%%%%%%
\begin{figure}[t]
 \begin{center}
  \includegraphics[width=0.45\textwidth]{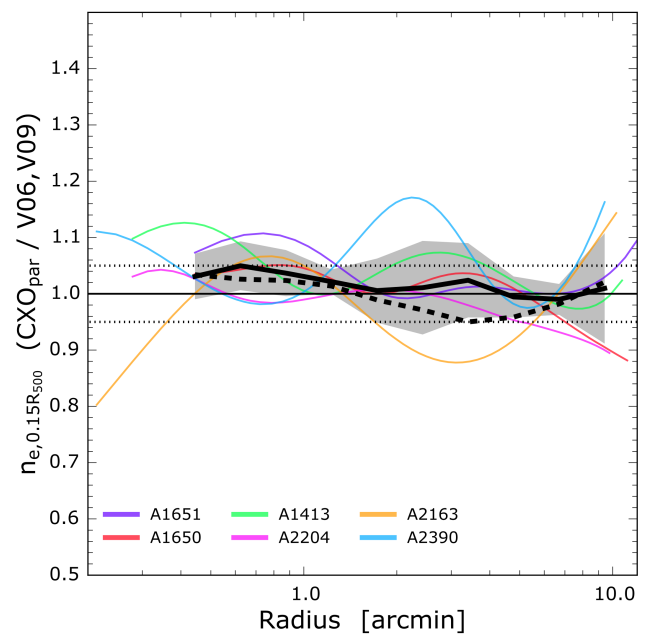}
 \end{center}
 \caption{\footnotesize Test of the \chandra\ photon weighting implementation developed in the present paper (Sect.~\ref{sec:weight} and Appendix~\ref{appendix_weight}). The Figure shows the ratio between the parametric \chandra\ density profiles derived with the weighting method developed here (pre-correction) and those published in V06/V09 (post-correction). Profiles are scaled by $n_{\rm e,0.15\,R_{500}}$ to compare the shape.  Legend is the same as in \figiac{fig:cons_dens}. As for \figiac{fig:cons_dens}, deviations in the core ($r < 0.5'$) are due to the different centres used for profile extraction.}
 \label{fig:ne_parvspar_vik}
\end{figure}
%%%%%%%%%%%%%%%%%%%%%%%%%%%%%%%%%%%%%%%%%%%%%%%%%%%%%%%%%%%%%%%%%%%%%%%%%%

\subsection{First Comparison }\label{first_comp}

Figure~\ref{fig:cons_dens} presents the ratio between the deprojected \xmm\ density profiles obtained using the \rexcess\ method and published by \citet[][]{PERxmmesz}, and the \chandra\ parametric profiles obtained using the \cccp\ method and published in V06. These are completely independent measurements, both in terms of instruments and methods.  We recall that the \rexcess\ analysis is  based on photon weighting for vignetting correction, and deprojection of the surface brightness profiles.  In the \cccp\ analysis,  the emission measure profiles, obtained after conversion of the observed count rate using the instrument response at each radius, are fitted with a general parametric model. The excluded point sources, the  profile extraction centre, and substructure masking are those of the original published works \citep[][V06/V09]{PERxmmesz}, and are thus also independent. For a correct comparison when comparing the \xmm\ profiles to V06/V09, we mask the contribution from the inner $\SI{40}{\kilo\parsec}$ as was done in these  works.
  
This first comparison provides an initial test of the robustness of the density profile reconstruction in general, and is representative of the maximum variation we can expect using different instruments, data, and profile extraction analysis.
The overall consistency  between the profiles is good: as shown by the shaded envelope, they agree to better than $5\%$ on average across the full radial range. The largest outlier is A2163, a complex object for which profile reconstruction is particularly sensitive to the masking of substructures and the choice of centre, which are different in the two works we compare here. 
%We notice a difference in the very inner parts, which is simply due to different centres used for profile extraction.
 Above $0.1R_{500}$ the profiles have similar forms, the ratio being almost constant. The median ratio suggests a small normalisation offset, the \chandra\ profiles being ~$\sim 2\%$ higher than \xmm,  although both instruments give consistent measurements within the errors.
We conclude that such completely independent analyses, using very different methods, may result in differences in absolute density values of the order of a few percent.

We will now proceed to undertake a number of further tests designed to investigate the impact on the density profiles of various data treatment choices and assumptions.

%%%%%%%%%%%%%%%%%%%%%%%%%%%%%%%%%%%%%%%%%%%%%%%%%%%%%%%%%%%%%%%%%%%%%%%%%%
\begin{figure*}[t]
\begin{center}
\includegraphics[width=0.9\textwidth]{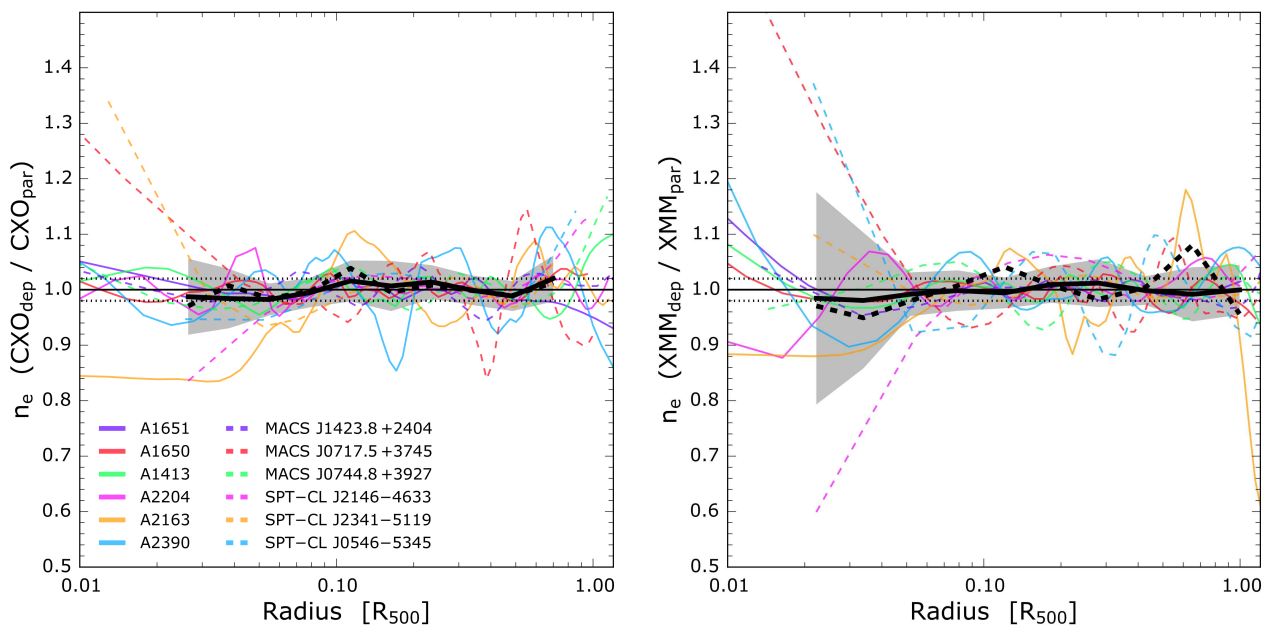} 
\caption{\footnotesize Test of parametric versus deprojection methods. Left panel: ratio between the \chandra\ density profiles obtained using the deprojection and parametric fit techniques. All other aspects of the analysis are identical. The legend is the same as in \figiac{fig:cons_dens} except that the dotted lines represent $\pm 2\%$ levels here. Right panel: same as in the left panel, but for the \xmm\ density profiles.}
\label{fig:cxo_xmm_decpar}
\end{center}
\end{figure*}
%%%%%%%%%%%%%%%%%%%%%%%%%%%%%%%%%%%%%%%%%%%%%%%%%%%%%%%%%%%%%%%%%%%%%%%%%%

\subsection{Tests of effective area correction method}\label{par_vs_vig}

As described in \seciac{sec:weight}, the effective area correction can be performed after profile extraction using the local effective area (post-correction),  or by weighting the photons before profile extraction (pre-correction). To investigate the impact of the different methods, and to validate the weighting procedure we developed here for \chandra\ data,   we compare the \chandra\ parametric density profiles derived with the weighting method for the six nearby clusters to those published in V06/V09. To account for possible effects introduced by using different versions of the {\tt CALDB}, we normalised the profiles by the density computed at $0.15\,R_{500}$. Figure \ref{fig:ne_parvspar_vik} shows that the agreement in profile shapes is remarkably good: the median ratio has maximum variations of the order of $\sim 1\%$ around unity.
As in \figiac{fig:cons_dens}, the profiles differ in the centre, sometimes by large amounts, simply because different centres have been used for the profile extraction. 
The absence of evident biases or trends with increasing off-axis angle implies that the weighting correction technique does not introduce any systematic uncertainties into the analysis.
There is a large dispersion in the $\sim 8-9$ arcmin region. This is due to the fact that most of the annuli in this region reach the border of the ACIS chips, where we expect poor count rates and hence noisy measurements.
Given the good agreement discussed above, in the following, all further tests will be undertaken on profiles extracted with the photon weighting (pre-correction) method.

\subsection{Tests of parametric versus non-parametric methods}\label{par_vs_nonpar}

In this Section we test the robustness of the reconstruction to the method used to derive  (3D) density profiles from the projected data.  Figure \ref{fig:cxo_xmm_decpar} shows the comparison between the density profiles derived using the deprojection method  to those derived from  parametric fitting. The ratios are  shown for the \chandra\ and \xmm\  sample in the left and right panels, respectively.
In both cases there is an excellent agreement above $0.02\, R_{500}$,  the median value being close to one, and the uncertainty envelope being of the order of $\sim 1\%$. There are strong deviations in the inner parts of the \xmm\ sample and a larger scatter in the central parts of both datasets. This is probably related to the presence of disturbed clusters, for which the parametric model has a tendency to smooth the variations of the surface brightness profiles. In the following, all further tests will be undertaken on profiles extracted with the non-parametric (deprojection) method.

%%%%%%%%%%%%%%%%%%%%%%%%%%%%%%%%%%%%%%%%%%%%%%%%%%%%%%%%%%%%%%%%%%%%%%%%%%
  \begin{figure}[t]
 \begin{center}
  \includegraphics[width=0.45\textwidth]{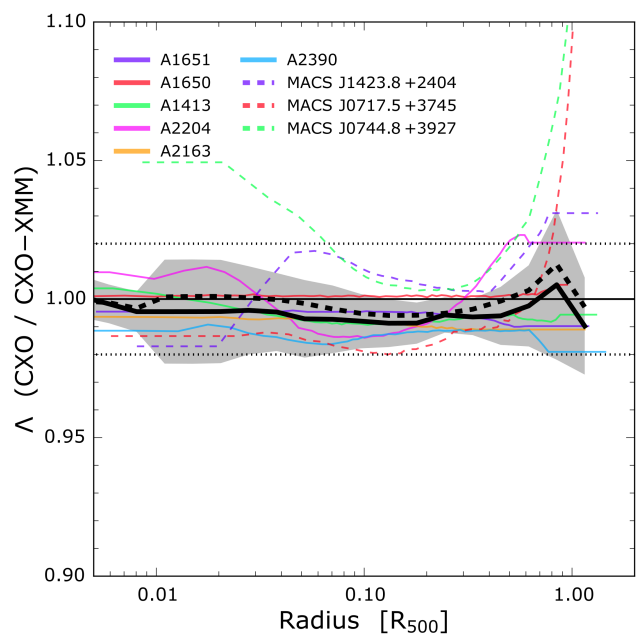}
 \end{center}
 \caption{{\footnotesize Effect of systematics on temperature profiles.   Ratio of $\Lambda(T)$ profiles derived for \chandra\  data,  using the temperature profiles measured by \chandra\ and \xmm. The $z>0.9$ clusters are excluded as the \chandra\ observations are not sufficiently deep for temperature profiles to be measured. 
 The legend is the same as in \figiac{fig:cons_dens} except that horizontal dotted lines represent $\pm 2\%$ levels.}}
 \label{fig:all_lambda}
\end{figure}
%%%%%%%%%%%%%%%%%%%%%%%%%%%%%%%%%%%%%%%%%%%%%%%%%%%%%%%%%%%%%%%%%%%%%%%%%%

%%%%%%%%%%%%%%%%%%%%%%%%%%%%%%%%%%%%%%%%%%%%%%%%%%%%%%%%%%%%%%%%%%%%%%%%%%
\begin{figure}[t]
 \begin{center}
  \includegraphics[width=0.45\textwidth]{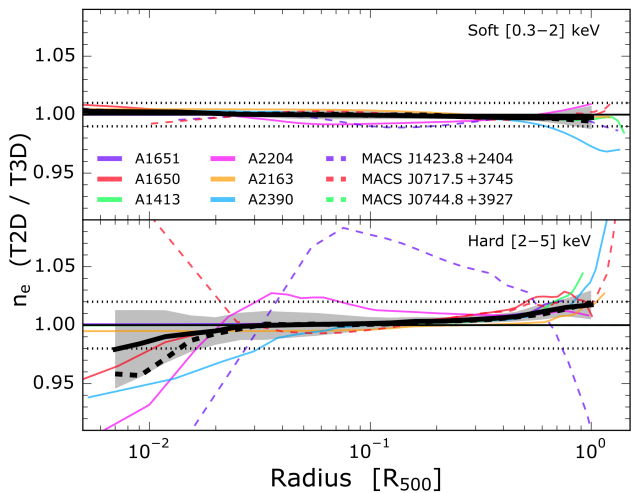}
 \end{center}
 \caption{\footnotesize Ratio between the density profiles obtained using  2D and 3D temperature profiles. {\it Top panel:} The density profiles are derived from the \xmm\ surface brightness extracted in the $[0.3-2]$ keV energy band.  The effect of using 2D profiles is negligible. The three highest redshift clusters, shown with dotted lines, are not included in the computation of the median/mean and its dispersion (see text for details).  {\it Bottom panel:} Profiles are obtained from surface brightness profiles extracted in the $[2-5]$ keV band. Here it is important to use 3D  profiles since the surface brightness is particularly sensitive to the temperature. The legend is the same as in \figiac{fig:cons_dens} except that horizontal dotted lines represent the $\pm 2\%$ levels.  As in the upper panel, the three highest redshift clusters are not included in the computation.}
 \label{fig:t2dt3d}
\end{figure} 
%%%%%%%%%%%%%%%%%%%%%%%%%%%%%%%%%%%%%%%%%%%%%%%%%%%%%%%%%%%%%%%%%%%%%%%%%%

\subsection{Impact of temperature profiles}\label{lambda_study}

As explained in \seciac{emission_measure}, density profile reconstruction 
relies on the cooling function, $\Lambda(T)$, used to convert count rates to emission measure profiles. We first investigate  the impact of the offset between \chandra\ and \xmm\ temperature measurements.
We expect the effect to be small, as the cooling function depends only weakly on temperature at the typical energies used for surface brightness extraction, $\sim [0.5-2]$ keV.
For each \chandra\ profile, we computed the $\Lambda(T)$ profile using the \xmm\ temperature profiles, namely $\Lambda_{\rm CXO-XMM}$. In \figiac{fig:all_lambda},  these profiles are compared  to the nominal profiles computed from  the \chandra\ temperature profile, $\Lambda_{\rm CXO}$. Over the full radial range the profiles show an agreement within $1\%$. 
Comparing density profiles reconstructed using $\Lambda_{\rm CXO}$ and $\Lambda_{\rm CXO-XMM}$ yields differences of the order of $<0.5\%$. Thus the offset in temperature measurements between \xmm\ and \chandra\ does not affect density profile reconstruction.

The $\Lambda(T)$ function is nominally computed using the projected (2D) temperature profile. In principle the emissivity should be computed in each shell, i.e. using the 3D temperature profile. It may differ from the 2D profile in the case where there are large temperature gradients, such as in the centres of cool core clusters. We derived the 3D  temperature profile from a parametric fit of the 2D profile, using the weighting scheme introduced by \citet{mazzotta2004} and  \citet{vikh_multit} to correct for spectroscopic bias. The average temperature in each annulus was computed from the contribution of each shell, taking into account the projection and PSF geometrical factors, and applying a density- and temperature-dependent weight, as defined by  \citet{vikh_multit}. 
This process  is in principle iterative, as the 3D temperature profile depends on the density profiles via the weights, while the density depends on the temperature via the cooling function.  In practice, we did not iterate the process, the final correction being very small. 

%%%%%%%%%%%%%%%%%%%%%%%%%%%%%%%%%%%%%%%%%%%%%%%%%%%%%%%%%%%%%%%%%%%%%%%%%%
\begin{figure}[t]
 \begin{center}
  \includegraphics[width=0.45\textwidth]{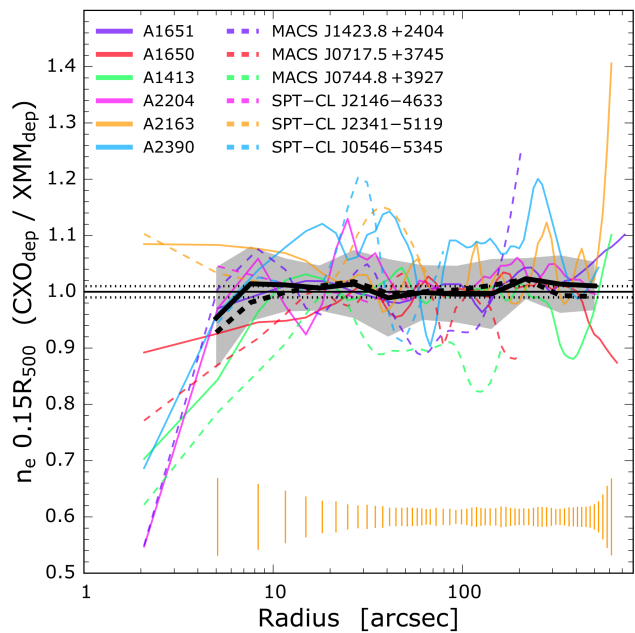}
 \end{center}
 \caption{\footnotesize Test of the \xmm\ PSF correction in the soft energy band: ratio between normalised density profiles obtained from the deprojection of  \chandra\ and \xmm\ surface brightness profiles. The error bars correspond to the error on the A2163 profile, and represent the typical uncertainties as a function of radius. Each profile is normalized by the  density computed at $R=0.15R_{500}$ to assess shape differences. There is an excellent agreement between the profile  shape above $5\arcsec$, showing that the \xmm\ PSF is properly accounted for in the density reconstruction. Legend is the same as in \figiac{fig:cons_dens} except that horizontal dotted lines represent $\pm 1\%$ levels.}
 \label{fig:xmm_ne_psftest}
\end{figure}
%%%%%%%%%%%%%%%%%%%%%%%%%%%%%%%%%%%%%%%%%%%%%%%%%%%%%%%%%%%%%%%%%%%%%%%%%%
To study the impact of using 2D rather than 3D temperature profiles, we examine the ratio of the density profiles computed with the corresponding 2D and 3D cooling functions. The density profiles are generally derived from the soft band surface brightness because of the higher S/N and the decreasing temperature dependence of the emissivity with energy. However, the emissivity depends on the {\it effective} temperature $kT_{\rm eff}=kT/(1+z)$, so that the soft band in fact behaves like the hard band for sufficiently high z clusters (e.g.  $kT_{\rm eff}  < 2$ kev). In \figiac{fig:t2dt3d} we thus show the profile ratios obtained in both the soft ($[0.3-2]$ keV) and hard ($[2-5]$ keV) energy bands. 
As expected, the density profiles of the low redshift sample, derived from soft energy band data, are perfectly consistent, well within $1\%$, as the dependency of the emissivity on the temperature is extremely weak at low energies. 
 For the three clusters at $z>0.25$ with temperature profiles,  the effective temperature remains above $2$ keV, and there is indeed good agreement between the 2D and the 3D results. However, with increasing redshifts the spectrum is shifted to lower energies, and the exponential cut off may reach the soft band used to extract the profiles and a full 3D analysis may become necessary. Unfortunately, for the three clusters at $z\sim1$ the observation depth is insufficient to determine their temperature profiles so that this point cannot  directly be assessed.
 We did find significant significant differences in the hard band, that become important for clusters with a complex morphology or strong cool core clusters (such as A2204). For this reason, one needs to use the 3D temperature profiles when deprojecting hard band surface brightness, or high $z$ clusters, particularly at low temperature.

\subsection{Tests of the \xmm\ PSF correction}

One crucial point of the density profile reconstruction is the correct estimation of the $[R_{\rm PSF}]$ term and the corresponding correction of the PSF effect.  This effect is only important for \xmm, for which the PSF\footnote{Values taken from the \xmm\ User's Handbook.} has a full width at half-maximum ${\rm FWHM} \sim 6\arcsec$ and a half-energy-width ${\rm HEW} \sim 15\arcsec$. This PSF size becomes particularly important for high redshift clusters. Typical $R_{500}$ values for a $z \sim 1$ cluster are of the order of $\sim 1-2$ arcmin; a resolution of $6\arcsec$ at the same redshift corresponds to 56~kpc for the cosmology we assume here. We thus exploited the excellent resolution of \chandra\ to undertake a set of tests to probe the robustness of the \xmm\ PSF model.

\subsubsection{Test of the \xmm\ PSF correction at low energy}\label{psf_study}

As a first test, we investigated if the PSF is properly corrected for in the $[0.3-2]$ keV energy range we use to extract the \xmm\ surface brightness profiles. In \figiac{fig:xmm_ne_psftest} we compare the \chandra\ and \xmm\ density profiles rescaled by the respective $n_{e}$ computed at $0.15\,R_{500}$. 
Above $\sim 6\arcsec$ the profiles are in excellent agreement, the median values scattering around unity with oscillations of the order of $0.5\%$. That is, in the energy range under consideration, the $[R_{\rm PSF}]$ term accurately reproduces the behaviour of the \xmm\ PSF. Density profile reconstruction is thus robust to the \xmm\ PSF down to the resolution of its typical FWHM. For a cluster of $M_{500} = 4 \times 10^{14}$ M$_{\odot}$ at $z=1$, this resolution limit corresponds to typical scales of $R/R_{500} \sim 0.07$. Below $\sim 6\arcsec$, the dispersion increases and the \xmm\ profiles appear too peaked on average. Here we are entering the very core of the \xmm\ PSF, where the spatial information is lost.  
%%%%%%%%%%%%%%%%%%%%%%%%%%%%%%%%%%%%%%%%%%%%%%%%%%%%%%%%%%%%%%%%%%%%%%%%%%
\begin{figure}[t]
 \begin{center}
  \includegraphics[width=0.45\textwidth]{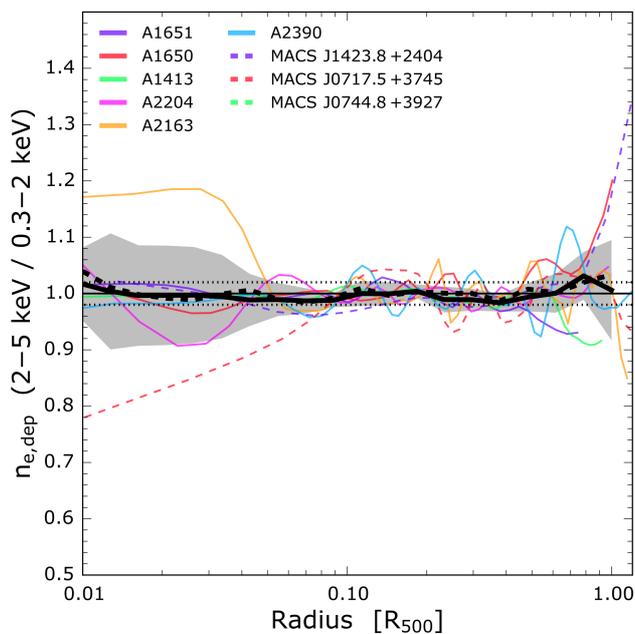} 
 \end{center}
 \caption{\footnotesize Test of the \xmm\ PSF correction in the hard energy band. Comparison between the deprojected density profile obtained from the \xmm\ surface brightness profiles extracted in the $[0.3-2]$ keV and in the $[2-5]$ keV band. Legend is the same as in \figiac{fig:cons_dens} except that dotted lines represent $\pm 2\%$ level.}
 \label{fig:psf_xmm_energy}
\end{figure}
%%%%%%%%%%%%%%%%%%%%%%%%%%%%%%%%%%%%%%%%%%%%%%%%%%%%%%%%%%%%%%%%%%%%%%%%%%

 %%%%%%%%%%%%%%%%%%%%%%%%%%%%%%%%%%%%%%%%%%%%%%%%%%%%%%%%%%%%%%%%%%%%%%%%%%
 \begin{figure}[t]
 \begin{center}
 \includegraphics[width=0.48\textwidth]{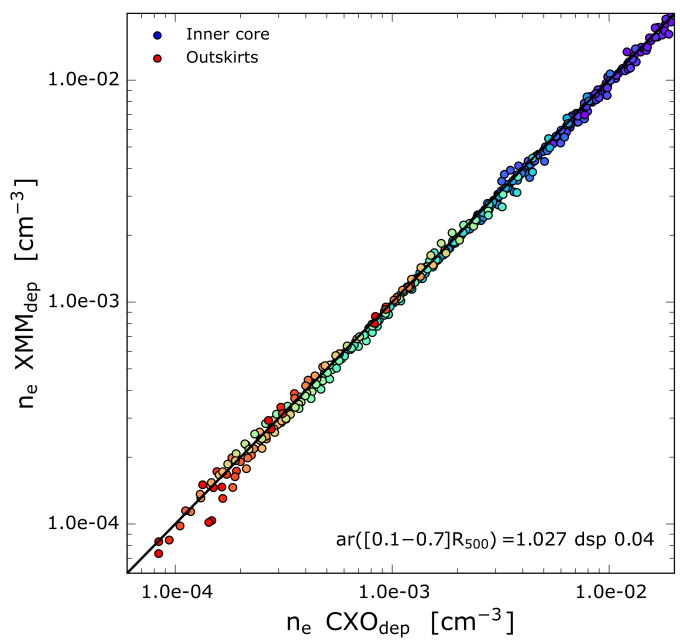}
  \end{center}
\caption{\footnotesize Deprojected \xmm\ and \chandra\ densities. Points are color coded according to the rainbow table to help 
the reader identify the inner and outer regions of the cluster, plotted with red and blue, respectively. The black solid line represents the identity relation. The average ratio between all the profiles in the $[0.1-0.7]\, R_{500}$ region is $1.027$ with a dispersion (dsp) of $0.04$.}
 \label{fig:brute_cxo_vs_xmm}
\end{figure}
%%%%%%%%%%%%%%%%%%%%%%%%%%%%%%%%%%%%%%%%%%%%%%%%%%%%%%%%%%%%%%%%%%%%%%%%%%

\subsubsection{Test of the energy dependence of the \xmm\ PSF}\label{psf_comparison}
As a by-product of this analysis we were also able to investigate whether  the \xmm\ PSF correction is  also accurate at high energy. Knowledge of the energy dependence is crucial for derivation of the temperature profile, the temperature measurement relying on the full energy band. For instance, incomplete knowledge of the \xmm\ PSF has been suggested as an explanation for the difference in the temperature profile shape obtained with \xmm\ and \chandra\ by \citet{donahue2014}.
Figure \ref{fig:psf_xmm_energy} shows the comparison between the density profiles obtained using the hard and soft band surface brightness, where for the hard band we use 3D temperature profiles to compute the cooling function (see Sec.~\ref{lambda_study} above). The agreement is remarkably good, the median ratio profile presenting oscillations of maximum $2\%$ around unity. There is a large dispersion in the central regions, which is driven by the presence of clusters with a complex morphology such as A2163 and MACS\,J0717.  However, above $0.1\,R_{500}$  the $1\sigma$ dispersion remains within $\pm 10\%$ (and within $\pm2\%$ below 0.6$R_{500}$). From this test we conclude that the model of \xmm\ PSF also correctly reproduces the correct  behaviour as a function of energy.

%%%%%%%%%%%%%%%%%%%%%%%%%%%%%%%%%%%%%%%%%%%%%%%%%%%%%%%%%%%%%%%%%%%%%%%%%%
\begin{figure*}
 \begin{center}
   \includegraphics[width=0.9\textwidth]{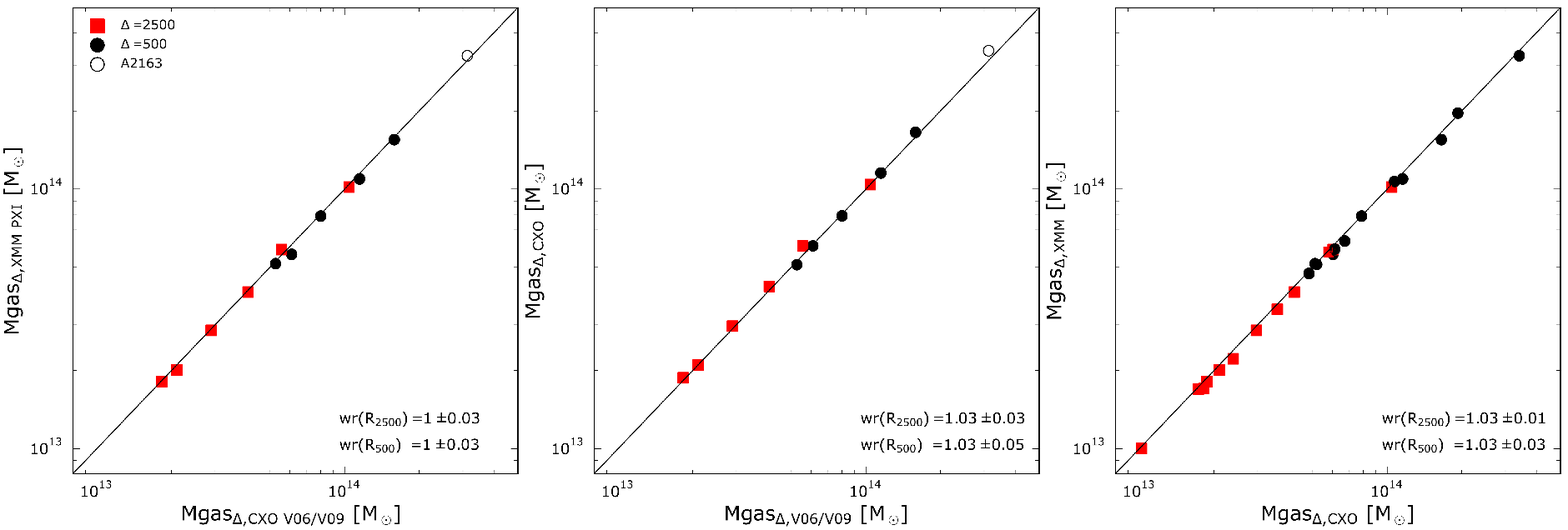}
 \end{center}
 \caption{\footnotesize Ratios between the integrated gas masses within $R_{2500}$ and $R_{500}$ obtained from the \xmm\ and the \chandra\ data. {\it For all panels:} the solid line represents the identity relation. {\it Left panel:} \xmm\ gas masses obtained by \citet{PERxmmesz} (PXI) versus \chandra\ gas masses from  \citet{vikhlinin2006,vikh2009}. {\it Central panel:} \chandra\ gas masses derived from the  pre-correction (photon weighting) method compared to those published in V06/V09.  {\it Right panel:}  \xmm\ and \chandra\ gas masses  obtained using the same analysis method (photon weighting method and non-parametric deconvolution/deprojection of the surface brightness profiles. }
 \label{fig:total_gas_mass}
\end{figure*}
%%%%%%%%%%%%%%%%%%%%%%%%%%%%%%%%%%%%%%%%%%%%%%%%%%%%%%%%%%%%%%%%%%%%%%%%%%

 \subsection{Absolute density comparison} \label{xmm_vs_cxo}

As already shown in \figiac{fig:xmm_ne_psftest}, the agreement in density profile shape between the instruments is excellent. We now focus on the flux differences, i.e. differences in terms of absolute normalisation. Figure \ref{fig:brute_cxo_vs_xmm} shows the scatter plot between all the data points from the \chandra\ and \xmm\ profiles.  The analysis method is the same for both observatories, and is based on the photon  weighting method and non-parametric deprojection/deconvolution of the surface brightness profiles. 
We colour-code the points following the rainbow table to clearly identify the inner, higher density (blue points), and the outer, lower density (red points) regions.
Focussing on the $[0.1-0.7]\, R_{500}$ region, we clearly see the presence of an offset, with the average ratio of all the values in this region yielding $1.027$ with a dispersion of $0.04\%$. This value is in agreement with previous work by \citet{martino2014} and \citet{donahue2014}, and is consistent with unity over the radial range under consideration.

Inspection of individual cluster profiles in \figiac{fig:neall} reveals that the normalisation offset presents complex behaviour. While A2204, A1413, A2390, and the three MACS clusters present a clear offset, the other clusters show good agreement, including in terms of normalisation. The normalisation offset thus seems to depend also on individual cluster observations, with a general trend for \chandra\ fluxes being slightly higher. This may  reflect a  dependence on observation conditions (e.g. residual soft proton contamination), or on observing epoch (if the calibration does not follow perfectly the instrument evolution).  It may also depend on intrinsic cluster properties, e.g. a physical effect  such as clumpiness, whose impact on the final density value may depend on the instrument sensitivity in the extraction band. 
 
%============================================================================================================================================================
\section{Consequences for the gas mass}

Gas mass is a fundamental quantity used to characterise galaxy clusters. Together with the stellar mass it yields the total amount of baryons, and if the total cluster mass is known, one can calculate the baryon fraction. The latter can be used for cosmological studies \citep[see e.g.][]{sasaki1996, pen1997,vikh_cosmo,mantz2014}. The quantity \YX, the product of the gas mass and the temperature  introduced by  \citet{krav2006},  is a low scatter proxy heavily used for the total cluster mass. It is the analogue of the integrated Compton parameter $Y$, thus linking X-ray and Microwave based observations. For these reasons, it is fundamental to investigate the impact of different analysis techniques, and of the normalisation offset, on the gas mass profile reconstruction and the total gas masses derived from these. In particular, $M_{\rm gas, \Delta}$ computed at $\Delta=2500,500$ are of notable importance\footnote{$R_{2500} \approx 0.45 R_{500}$.}. 

Figure \ref{fig:total_gas_mass} shows the gas mass  computed at $R=R_{2500}$ and $R=R_{500}$ using red squares and black dots, respectively. We show three comparisons,  corresponding to the tests performed in  \seciac{first_comp}, \seciac{par_vs_vig}, and \seciac{xmm_vs_cxo} respectively, viz.,

\begin{itemize}
\item The left panel shows the scatter plot between the gas masses  independently derived by \citet{PERxmmesz}  and by \citet{vikhlinin2006,vikh2009}. We recall that the former are derived from \xmm\ data using the pre-correction (photon weighting) method and using deprojection/deconvolution of the surface brightness profiles. The latter are  obtained  from \chandra\ data using the post-correction method and parametric gas density profile models. There is  remarkably good agreement at  $R_{2500}$, the weighted ratio of this subsample yielding ${\rm WR}_{2500}=1.00 \pm 0.03$. The agreement is also good when considering the $R_{500}$ subsample. The corresponding weighted ratio is  ${\rm WR}_{500}=1.00 \pm 0.03$.
\item The central panel tests the impact of the  vignetting correction method. The \chandra\ gas masses obtained using the pre-correction (photon weighting) method  are compared to the values published in V06/V09. 
There is no evident trend or behaviour, as all the points are aligned with the 1:1 slope.  Computing a weighted ratio at $R_{2500}$ yields ${\rm WR}_{2500}=1.03 \pm 0.03$. This value confirms the presence of the offset, due to the different calibration files used between the V06/V09 analysis and the analysis presented in this paper. 
Removing $A2163$ does not change the result. At $R_{500}$ the weighted ratio is ${WR}_{500}=1.03 \pm 0.03$ considering $A2163$; removing this cluster yields ${\rm WR}_{500}=1.01 \pm 0.03$. This result again confirms the excellent agreement between measurements performed using completely independent analyses.
\item Finally, the right panel compares the \xmm\ and \chandra\ gas masses obtained using the same analysis method (pre-correction and non-parametric deprojection, in both cases).
  Both subsamples show the same behaviour, all the points being slightly below the unity line. 
  This result is confirmed by the weighted ratios ${\rm WR}_{2500}=1.03 \pm 0.01$ and ${\rm WR}_{500}=1.03 \pm 0.03$. In this case A2163 is less problematic because  in this comparison the X-ray peaks on which we centre our profiles are the same. \chandra\ gas masses are  slightly higher, though consistent, with those from \xmm.
\end{itemize}

%%%%%%%%%%%%%%%%%%%%%%%%%%%%%%%%%%%%%%%%%%%%%%%%%%%%%%%%%%%%%%%%%%%%%%%%%%
\section{Discussion and conclusions}\label{conclusions}

In this work we have compared the gas density profiles and integrated gas masses  obtained from \chandra\ and \xmm\ observations for a sample of twelve  clusters. We have undertaken a thorough investigation of potential sources of systematic differences between measurements, taking into account that the data originate from different observatories, and examining subtle effects linked to the analysis method used to reconstruct the gas density.  We summarise the results of each test in \seciac{appendix_summary}.

Our main conclusions are as follows:

\begin{itemize}
\item Direct comparison of previously-published results, obtained from \chandra\ for the \cccp\  project \citep{vikhlinin2006,vikh2009} and from \xmm\ for \planck\ studies \citep{PERxmmesz},  yield excellent agreement. The differences for individual objects remain within $\pm5\%$ at any radius, except for very specific features, and there are no obvious systematic trends. This is remarkable, given the different instruments, treatment of instrumental effects such as vignetting,  deprojection methods (parametric versus non parametric), and even the point source masking, choice of centre, background subtraction etc..  
\item For \chandra, we implement and validate  the event weighting procedure to correct for vignetting before profile extraction, finding similar results to methods that correct a posteriori the surface brightness profiles using the local effective area. 
\item Density profile reconstruction using parametric fitting of projected profiles (emission measure or surface brightness), or deconvolution/deprojection techniques, show on average an exquisite agreement (better than 1\% on average).  However, individual fluctuations at various radii can reach up to $10\%$, linked to the different sensitivity of the reconstruction techniques to small scale surface brightness features, and to the underlying dynamical state of the object in question. 
\item There is a nearly perfect consistency of the shapes of  \chandra\ and \xmm\ density profiles beyond $\sim 6\arcsec$ (i.e. the very core of the \xmm\ PSF). 
Our understanding of the \xmm\ PSF is thus excellent. Further checks show that we are also able to accurately correct for the PSF at high energy. Insufficient knowledge of the \xmm\ PSF was suggested by  \citet{donahue2014} to at least partly account for the differences in the radial temperature profiles they derived from  \xmm\ and \chandra. Our result does not support this view.
\item The known temperature offset between \chandra\ and \xmm\ has almost no impact (less than $0.5\%$) on the resulting density profile. The approximation made by using the projected (2D) temperature profile to convert count rate to density is negligible in the soft energy band, even in the case of strong temperature gradients, e.g. cool cores. However when working in the hard band, or equivalently in the soft band but for very high redshift systems, it is essential to account for the deprojected  radial temperature profile. 
\item The overall normalisation offset remains within $\sim 2.5\%$.\footnote{On average we find a small, though not significant at the $1\sigma $ level,  offset between the absolute value of the  \chandra\ and \xmm\ density profiles, the former being higher on average by $1.027$ with a dispersion of $0.04$.} This effect is dominated by the  absolute flux calibration, which can depend on the calibration data base version. A second order effect is the dependence on individual  observations, linked to cluster properties and/or observation conditions and/or instrument evolution. 

\item Gas mas profiles generally follow the same behaviour as the density. Gas masses at $R_{500}$ or $R_{2500}$ present a small offset,  but on the average the measurements are consistent within $1\sigma$.
\end{itemize}

Our analysis confirms the good agreement  generally found in the literature \citep[e.g.,][]{martino2014,donahue2014},  for comparisons of density profiles obtained with \xmm\ and \chandra\ on specific samples analysed with a given procedure.   Our study provides a more complete understanding of the different sources of systematic effects, across the full cluster population.  The present sample was specifically chosen to cover a wide redshift range (from the local universe to $z\sim1$), and to sample the variety of dynamical  states (from relaxed objects to violent mergers).  We address both instrumental effects, by comparing between satellites,  and the effect of differing analysis methods, by comparing results obtained from the same satellite and varying the analysis.  We emphasize that, on average, the density profiles are robust to the analysis method (e.g. vignetting correction, parametric versus non parametric modeling of the surface brightness); that the effect of the \xmm\ PSF is well understood and can be corrected at small radii deep into the cluster core; and that the overall density normalisation offset remains within $\sim2.5\%$.  The results presented here are  important in the context of any project that envisions the combination of  \chandra\ and \xmm\ data. Indeed, we  would be foolhardy not to do so, given the  complementarity that such observations afford.

%-------------------------acknowledgements-------------------------------------------------------------------------------------------
\begin{acknowledgements} 
The authors thank the referee for his/her comments.
The scientific results reported in this article are based on data obtained from the \chandra\ Data Archive and observations obtained with \xmm , an ESA science mission with instruments and contributions directly funded by ESA Member States and NASA.
 The research leading to these results has received funding from the European  Research  Council  under  the  European  Union’s  Seventh  Framework
Programme (FP72007-2013) ERC grant agreement no 340519. M.A., P.M., E.P., and GWP acknowledge partial funding support from NASA grant NNX14AC22G. F.A-S. acknowledges support from \chandra\ grant GO3-14131X.
 \end{acknowledgements}
%-------------------------acknowledgements-------------------------------------------------------------------------------------------

\bibliographystyle{aa}
\bibliography{lib_density_paper}

\begin{thebibliography}{47}
\expandafter\ifx\csname natexlab\endcsname\relax\def\natexlab#1{#1}\fi

\bibitem[{{Allen} {et~al.}(2011){Allen}, {Evrard}, \& {Mantz}}]{allen2011}
{Allen}, S.~W., {Evrard}, A.~E., \& {Mantz}, A.~B. 2011, \araa, 49, 409

\bibitem[{{Arnaud} {et~al.}(2001){Arnaud}, {Neumann}, {Aghanim}, {Gastaud},
  {Majerowicz}, \& {Hughes}}]{arnaud2001}
{Arnaud}, M., {Neumann}, D.~M., {Aghanim}, N., {et~al.} 2001, \aap, 365, L80

\bibitem[{{Arnaud} {et~al.}(2010){Arnaud}, {Pratt}, {Piffaretti},
  {B{\"o}hringer}, {Croston}, \& {Pointecouteau}}]{arnaud2010}
{Arnaud}, M., {Pratt}, G.~W., {Piffaretti}, R., {et~al.} 2010, \aap, 517, A92

\bibitem[{{Bartalucci} {et~al.}(2017){Bartalucci}, {Arnaud}, {Pratt},
  {D{\'e}mocl{\`e}s}, {van der Burg}, \& {Mazzotta}}]{bar17}
{Bartalucci}, I., {Arnaud}, M., {Pratt}, G.~W., {et~al.} 2017, \aap, 598, A61

\bibitem[{{Bartalucci} {et~al.}(2014){Bartalucci}, {Mazzotta}, {Bourdin}, \&
  {Vikhlinin}}]{bartalucci2014}
{Bartalucci}, I., {Mazzotta}, P., {Bourdin}, H., \& {Vikhlinin}, A. 2014, \aap,
  566, A25

\bibitem[{{Churazov} {et~al.}(2008){Churazov}, {Forman}, {Vikhlinin},
  {Tremaine}, {Gerhard}, \& {Jones}}]{churazov2008}
{Churazov}, E., {Forman}, W., {Vikhlinin}, A., {et~al.} 2008, \mnras, 388, 1062

\bibitem[{{Croston} {et~al.}(2006){Croston}, {Arnaud}, {Pointecouteau}, \&
  {Pratt}}]{croston2006}
{Croston}, J.~H., {Arnaud}, M., {Pointecouteau}, E., \& {Pratt}, G.~W. 2006,
  \aap, 459, 1007

\bibitem[{{Croston} {et~al.}(2008){Croston}, {Pratt}, {B{\"o}hringer},
  {Arnaud}, {Pointecouteau}, {Ponman}, {Sanderson}, {Temple}, {Bower}, \&
  {Donahue}}]{croston2008}
{Croston}, J.~H., {Pratt}, G.~W., {B{\"o}hringer}, H., {et~al.} 2008, \aap,
  487, 431

\bibitem[{{Donahue} {et~al.}(2014){Donahue}, {Voit}, {Mahdavi}, {Umetsu},
  {Ettori}, {Merten}, {Postman}, {Hoffer}, {Baldi}, {Coe}, {Czakon},
  {Bartelmann}, {Benitez}, {Bouwens}, {Bradley}, {Broadhurst}, {Ford},
  {Gastaldello}, {Grillo}, {Infante}, {Jouvel}, {Koekemoer}, {Kelson}, {Lahav},
  {Lemze}, {Medezinski}, {Melchior}, {Meneghetti}, {Molino}, {Moustakas},
  {Moustakas}, {Nonino}, {Rosati}, {Sayers}, {Seitz}, {Van der Wel}, {Zheng},
  \& {Zitrin}}]{donahue2014}
{Donahue}, M., {Voit}, G.~M., {Mahdavi}, A., {et~al.} 2014, \apj, 794, 136

\bibitem[{{Freeman} {et~al.}(2002){Freeman}, {Kashyap}, {Rosner}, \&
  {Lamb}}]{freeman2002}
{Freeman}, P.~E., {Kashyap}, V., {Rosner}, R., \& {Lamb}, D.~Q. 2002, \apjs,
  138, 185

\bibitem[{{Fruscione} {et~al.}(2006){Fruscione}, {McDowell}, {Allen},
  {Brickhouse}, {Burke}, {Davis}, {Durham}, {Elvis}, {Galle}, {Harris},
  {Huenemoerder}, {Houck}, {Ishibashi}, {Karovska}, {Nicastro}, {Noble},
  {Nowak}, {Primini}, {Siemiginowska}, {Smith}, \& {Wise}}]{fruscione2006}
{Fruscione}, A., {McDowell}, J.~C., {Allen}, G.~E., {et~al.} 2006, in
  \procspie, Vol. 6270, Society of Photo-Optical Instrumentation Engineers
  (SPIE) Conference Series, 62701V

\bibitem[{{Garmire} {et~al.}(2003){Garmire}, {Bautz}, {Ford}, {Nousek}, \&
  {Ricker}}]{garmire2003}
{Garmire}, G.~P., {Bautz}, M.~W., {Ford}, P.~G., {Nousek}, J.~A., \& {Ricker},
  Jr., G.~R. 2003, in \procspie, Vol. 4851, X-Ray and Gamma-Ray Telescopes and
  Instruments for Astronomy., ed. J.~E. {Truemper} \& H.~D. {Tananbaum}, 28--44

\bibitem[{{Ghizzardi}(2001)}]{ghizzardi2001}
{Ghizzardi}, S. 2001, XMM-SOC-CAL-TN-0022

\bibitem[{{Giacconi} {et~al.}(2001){Giacconi}, {Rosati}, {Tozzi}, {Nonino},
  {Hasinger}, {Norman}, {Bergeron}, {Borgani}, {Gilli}, {Gilmozzi}, \&
  {Zheng}}]{giacconi2001}
{Giacconi}, R., {Rosati}, P., {Tozzi}, P., {et~al.} 2001, \apj, 551, 624

\bibitem[{{Hickox} \& {Markevitch}(2006)}]{markevitch2006}
{Hickox}, R.~C. \& {Markevitch}, M. 2006, \apj, 645, 95

\bibitem[{{Kaastra}(1992)}]{kaastra1992}
{Kaastra}, J.~S. 1992, {An X-Ray Spectral Code for Optically Thin Plasmas
  (Internal SRON-Leiden Report, updated version 2.0)}

\bibitem[{{Kalberla} {et~al.}(2005){Kalberla}, {Burton}, {Hartmann}, {Arnal},
  {Bajaja}, {Morras}, \& {P{\"o}ppel}}]{kalberla2005}
{Kalberla}, P.~M.~W., {Burton}, W.~B., {Hartmann}, D., {et~al.} 2005, \aap,
  440, 775

\bibitem[{{Kravtsov} {et~al.}(2006){Kravtsov}, {Vikhlinin}, \&
  {Nagai}}]{krav2006}
{Kravtsov}, A.~V., {Vikhlinin}, A., \& {Nagai}, D. 2006, \apj, 650, 128

\bibitem[{{Kuntz} \& {Snowden}(2000)}]{kuntzsnowden2000}
{Kuntz}, K.~D. \& {Snowden}, S.~L. 2000, \apj, 543, 195

\bibitem[{{Liedahl} {et~al.}(1995){Liedahl}, {Osterheld}, \&
  {Goldstein}}]{liedhal1995}
{Liedahl}, D.~A., {Osterheld}, A.~L., \& {Goldstein}, W.~H. 1995, \apjl, 438,
  L115

\bibitem[{{Limousin} {et~al.}(2016){Limousin}, {Richard}, {Jullo}, {Jauzac},
  {Ebeling}, {Bonamigo}, {Alavi}, {Cl{\'e}ment}, {Giocoli}, {Kneib}, {Verdugo},
  {Natarajan}, {Siana}, {Atek}, \& {Rexroth}}]{limousin2016}
{Limousin}, M., {Richard}, J., {Jullo}, E., {et~al.} 2016, \aap, 588, A99

\bibitem[{{Lumb} {et~al.}(2002){Lumb}, {Warwick}, {Page}, \& {De
  Luca}}]{lumb2002}
{Lumb}, D.~H., {Warwick}, R.~S., {Page}, M., \& {De Luca}, A. 2002, \aap, 389,
  93

\bibitem[{{Mahdavi} {et~al.}(2013){Mahdavi}, {Hoekstra}, {Babul}, {Bildfell},
  {Jeltema}, \& {Henry}}]{mahdavi2013}
{Mahdavi}, A., {Hoekstra}, H., {Babul}, A., {et~al.} 2013, \apj, 767, 116

\bibitem[{{Mantz} {et~al.}(2014){Mantz}, {Allen}, {Morris}, {Rapetti},
  {Applegate}, {Kelly}, {von der Linden}, \& {Schmidt}}]{mantz2014}
{Mantz}, A.~B., {Allen}, S.~W., {Morris}, R.~G., {et~al.} 2014, \mnras, 440,
  2077

\bibitem[{{Markevitch}(2010)}]{chandra_back}
{Markevitch}, M. 2010, http://cxc.cfa.harvard.edu/contrib/maxim/

\bibitem[{{Martino} {et~al.}(2014){Martino}, {Mazzotta}, {Bourdin}, {Smith},
  {Bartalucci}, {Marrone}, {Finoguenov}, \& {Okabe}}]{martino2014}
{Martino}, R., {Mazzotta}, P., {Bourdin}, H., {et~al.} 2014, \mnras, 443, 2342

\bibitem[{{Mazzotta} {et~al.}(2004){Mazzotta}, {Rasia}, {Moscardini}, \&
  {Tormen}}]{mazzotta2004}
{Mazzotta}, P., {Rasia}, E., {Moscardini}, L., \& {Tormen}, G. 2004, \mnras,
  354, 10

\bibitem[{{Morrison} \& {McCammon}(1983)}]{morrison1983}
{Morrison}, R. \& {McCammon}, D. 1983, \apj, 270, 119

\bibitem[{{Parmar} {et~al.}(1999){Parmar}, {Guainazzi}, {Oosterbroek}, {Orr},
  {Favata}, {Lumb}, \& {Malizia}}]{parmar1999}
{Parmar}, A.~N., {Guainazzi}, M., {Oosterbroek}, T., {et~al.} 1999, \aap, 345,
  611

\bibitem[{{Pen}(1997)}]{pen1997}
{Pen}, U.-L. 1997, \na, 2, 309

\bibitem[{{Planck Collaboration Int. III}(2013)}]{PIPmass}
{Planck Collaboration Int. III}. 2013, \aap, 550, A129

\bibitem[{{Planck Collaboration XI}(2011)}]{PERxmmesz}
{Planck Collaboration XI}. 2011, \aap, 536, A11

\bibitem[{{Planck Collaboration XX}(2014)}]{P13szcount}
{Planck Collaboration XX}. 2014, \aap, 571, A20

\bibitem[{{Pratt} {et~al.}(2007){Pratt}, {B{\"o}hringer}, {Croston}, {Arnaud},
  {Borgani}, {Finoguenov}, \& {Temple}}]{pratt2007}
{Pratt}, G.~W., {B{\"o}hringer}, H., {Croston}, J.~H., {et~al.} 2007, \aap,
  461, 71

\bibitem[{{Pratt} {et~al.}(2009){Pratt}, {Croston}, {Arnaud}, \&
  {B{\"o}hringer}}]{pratt2009}
{Pratt}, G.~W., {Croston}, J.~H., {Arnaud}, M., \& {B{\"o}hringer}, H. 2009,
  \aap, 498, 361

\bibitem[{{Sasaki}(1996)}]{sasaki1996}
{Sasaki}, S. 1996, \pasj, 48, L119

\bibitem[{{Schellenberger} {et~al.}(2015){Schellenberger}, {Reiprich},
  {Lovisari}, {Nevalainen}, \& {David}}]{sch2015}
{Schellenberger}, G., {Reiprich}, T.~H., {Lovisari}, L., {Nevalainen}, J., \&
  {David}, L. 2015, \aap, 575, A30

\bibitem[{{Snowden} {et~al.}(1995){Snowden}, {Freyberg}, {Plucinsky},
  {Schmitt}, {Truemper}, {Voges}, {Edgar}, {McCammon}, \&
  {Sanders}}]{snowden1995}
{Snowden}, S.~L., {Freyberg}, M.~J., {Plucinsky}, P.~P., {et~al.} 1995, \apj,
  454, 643

\bibitem[{{Snowden} {et~al.}(2008){Snowden}, {Mushotzky}, {Kuntz}, \&
  {Davis}}]{snowden2008}
{Snowden}, S.~L., {Mushotzky}, R.~F., {Kuntz}, K.~D., \& {Davis}, D.~S. 2008,
  \aap, 478, 615

\bibitem[{Starck {et~al.}(1998)Starck, Murtagh, \& Bijaoui}]{sta98}
Starck, J.-L., Murtagh, F., \& Bijaoui, A. 1998, Image Processing and Data
  Analysis: The Multiscale Approach (New York, NY, USA: Cambridge University
  Press)

\bibitem[{{Str{\"u}der} {et~al.}(2001){Str{\"u}der}, {Briel}, {Dennerl},
  {Hartmann}, {Kendziorra}, {Meidinger}, {Pfeffermann}, {Reppin}, {Aschenbach},
  {Bornemann}, {Br{\"a}uninger}, {Burkert}, {Elender}, {Freyberg}, {Haberl},
  {Hartner}, {Heuschmann}, {Hippmann}, {Kastelic}, {Kemmer}, {Kettenring},
  {Kink}, {Krause}, {M{\"u}ller}, {Oppitz}, {Pietsch}, {Popp}, {Predehl},
  {Read}, {Stephan}, {St{\"o}tter}, {Tr{\"u}mper}, {Holl}, {Kemmer}, {Soltau},
  {St{\"o}tter}, {Weber}, {Weichert}, {von Zanthier}, {Carathanassis}, {Lutz},
  {Richter}, {Solc}, {B{\"o}ttcher}, {Kuster}, {Staubert}, {Abbey}, {Holland},
  {Turner}, {Balasini}, {Bignami}, {La Palombara}, {Villa}, {Buttler},
  {Gianini}, {Lain{\'e}}, {Lumb}, \& {Dhez}}]{struder2001}
{Str{\"u}der}, L., {Briel}, U., {Dennerl}, K., {et~al.} 2001, \aap, 365, L18

\bibitem[{{Turner} {et~al.}(2001){Turner}, {Abbey}, {Arnaud}, {Balasini},
  {Barbera}, {Belsole}, {Bennie}, {Bernard}, {Bignami}, {Boer}, {Briel},
  {Butler}, {Cara}, {Chabaud}, {Cole}, {Collura}, {Conte}, {Cros}, {Denby},
  {Dhez}, {Di Coco}, {Dowson}, {Ferrando}, {Ghizzardi}, {Gianotti}, {Goodall},
  {Gretton}, {Griffiths}, {Hainaut}, {Hochedez}, {Holland}, {Jourdain},
  {Kendziorra}, {Lagostina}, {Laine}, {La Palombara}, {Lortholary}, {Lumb},
  {Marty}, {Molendi}, {Pigot}, {Poindron}, {Pounds}, {Reeves}, {Reppin},
  {Rothenflug}, {Salvetat}, {Sauvageot}, {Schmitt}, {Sembay}, {Short},
  {Spragg}, {Stephen}, {Str{\"u}der}, {Tiengo}, {Trifoglio}, {Tr{\"u}mper},
  {Vercellone}, {Vigroux}, {Villa}, {Ward}, {Whitehead}, \&
  {Zonca}}]{turner2001}
{Turner}, M.~J.~L., {Abbey}, A., {Arnaud}, M., {et~al.} 2001, \aap, 365, L27

\bibitem[{{Vikhlinin}(2006)}]{vikh_multit}
{Vikhlinin}, A. 2006, \apj, 640, 710

\bibitem[{{Vikhlinin} {et~al.}(2009{\natexlab{a}}){Vikhlinin}, {Burenin},
  {Ebeling}, {Forman}, {Hornstrup}, {Jones}, {Kravtsov}, {Murray}, {Nagai},
  {Quintana}, \& {Voevodkin}}]{vikh2009}
{Vikhlinin}, A., {Burenin}, R.~A., {Ebeling}, H., {et~al.} 2009{\natexlab{a}},
  \apj, 692, 1033

\bibitem[{{Vikhlinin} {et~al.}(2006){Vikhlinin}, {Kravtsov}, {Forman}, {Jones},
  {Markevitch}, {Murray}, \& {Van Speybroeck}}]{vikhlinin2006}
{Vikhlinin}, A., {Kravtsov}, A., {Forman}, W., {et~al.} 2006, \apj, 640, 691

\bibitem[{{Vikhlinin} {et~al.}(2009{\natexlab{b}}){Vikhlinin}, {Kravtsov},
  {Burenin}, {Ebeling}, {Forman}, {Hornstrup}, {Jones}, {Murray}, {Nagai},
  {Quintana}, \& {Voevodkin}}]{vikh_cosmo}
{Vikhlinin}, A., {Kravtsov}, A.~V., {Burenin}, R.~A., {et~al.}
  2009{\natexlab{b}}, \apj, 692, 1060

\bibitem[{{Voit}(2005)}]{voit2005}
{Voit}, G.~M. 2005, Reviews of Modern Physics, 77, 207

\end{thebibliography}
\begingroup
\renewcommand{\cleardoublepage}{}
\renewcommand{\clearpage}{}
\appendix
\section{\chandra\ and \xmm\ data reduction}\label{appendix_reduction}
\endgroup
\subsection{\chandra\ data reduction}\label{chandra_data}
\chandra\ data are cleaned using the \chandra\ Interactive Analysis of Observations (CIAO, \citealt{fruscione2006}) tools version $4.7$ and the \chandra-ACIS calibration database version $4.6.5$ of December $2014$.  
We reprocess level $1$ event files using the \verb?chandra_repro? tool to apply the latest calibration files and produce new bad pixel maps.
Events which are likely to be due to high energetic particles are flagged using the Grade keyword and are removed from the analysis.
 In addition to the Grade filtering we also apply the Very Faint mode filtering which further reduces the particle contamination\footnote{cxc.harvard.edu/cal/Acis/Cal\_prods/vfbkgrnd}.

To remove periods affected by flare contamination we follow the procedures described in \cite{markevitch2006} and in the \chandra\ background COOKBOOK \cite{chandra_back}. For ACIS-I observations we extract the light curve from the four ACIS-I CCDs with a temporal bin of $\SI{259.28}{\second}$ in the $[0.3-12]\si{\kilo\electronvolt}$ band. For ACIS-S observations we extract the light-curve from CCD S3 with a bin of $\SI{1037.12}{\second}$ in the $[2.5-7]\si{\kilo\electronvolt}$ band. We then compute a mean value by fitting a Poisson distribution on the histogram of the lightcurve, using the \verb?lc_clean? script. We exclude from the analysis all time intervals where the count rate deviates more than $3\sigma$ from the mean value. 

%%%%%%%%%%%%%%%%%%%%%%%%%%%%%%%%%%%%%%%%%%%%%%%%%%%%%%%%%%%%%%%%%%%%%%%%%%
\subsection{\xmm\ data reduction}\label{xmm_data}
\xmm\ datasets are analysed using the Science Analysis System (SAS \footnote{cosmos.esa.int/web/xmm-newton}) version $15.0$ and the current calibration 
files as available to January $2017$.
Object data files are reprocessed to apply up-to-date calibration files, indexed by the \verb?cifbuild?, using the \verb?emchain? tool. From processed data we remove periods contaminated by flares following the procedures described in \cite{pratt2007}. We extract the lightcurves from each instrument and we use a Poisson curve to fit the light curve histogram and exclude periods where the count rate exceeds $3\sigma$. 

After the flare removal stage we apply a further cleaning filter based on the PATTERN keyword. Similar to the GRADE of \chandra\ this keyword is used to flag 
photons which are likely to be produced by the interaction of high energetic particles with the detector. Events whose PATTERN keyword is greater than $13$ and $4$, for the MOS$1-2$ and PN cameras, respectively, are removed from the analysis. After the cleaning stage, the three camera datasets are combined.

%%%%%%%%%%%%%%%%%%%%%%%%%%%%%%%%%%%%%%%%%%%%%%%%%%%%%%%%%%%%%%%%%%%%%%%%%%
\subsection{Point source masking}
We identify point sources in \chandra\ datasets running the CIAO wavelet detection tool \texttt{wavdetect} \citep{freeman2002} on $[0.5-1.2], [1.2-2]$ and $[2-7]$ keV exposure-corrected images. To detect point sources in \xmm\ datasets we run the Multiresolution wavelet software \citep{sta98} on the exposure-corrected $[0.3-2]$\, \si{\kilo\electronvolt} and $[2-5]$ keV images.
Both lists of regions are inspected by eye to check for false positives or missed sources and then merged, in order to remove the same regions from the analysis.

 In addition to the point sources, we mask the emission coming from the three sub-structures of 
MACSJ0717.5+3745, two in the South-East sector and one in the North-West (corresponding to the substructures identified as A,C, and D in \citealt{limousin2016}).

\subsection{Background estimation}\label{back_estimation}
\begin{table*}[h]
\caption{{\footnotesize Minimum and maximum values of the median of the profiles ratio (black solid line) for all the test presented in \seciac{robustness}. The values are expressed in term of $\%$ variation respect to 1.}}\label{tab:resume_table}
\begin{center}
\resizebox{\textwidth}{!} {
\begin{tabular}{lccc}
\hline        
\hline
Test  name      																																	&  	Section 							& Fig. associated							&	 Min/Max [$\%$]			\\
\hline
XMM (PX1) over CXO published in V06/V09																							&\seciac{first_comp}				& \figiac{fig:cons_dens}					&	-3 / +2					\\
Par. CXO $n_{e}$ over CXO $n_{e}$ published in V06/V09, normalised at $0.15R_{500}$ 						&\seciac{par_vs_vig}				& \figiac{fig:ne_parvspar_vik}			&	-1 / +5			\\
Dep. CXO $n_{e}$ over par. CXO $n_{e}$ 																							& \seciac{par_vs_nonpar}		& \figiac{fig:cxo_xmm_decpar}			&	-2 / +2	\\
Dep. XMM $n_{e}$ over par. XMM $n_{e}$ 																							& \seciac{par_vs_nonpar}		& \figiac{fig:cxo_xmm_decpar}			& -2 / +1	\\
CXO $\Lambda$ over CXO $\Lambda$ computed using XMM kT profiles 												&	\seciac{lambda_study}		& \figiac{fig:all_lambda}					&	-1 / +1	\\
XMM $n_{e}$  2D kT over XMM $n_{e}$  3D kT  in the $0.3-2$ keV band												&	\seciac{lambda_study}		& \figiac{fig:t2dt3d}							&	-0 / +0$^{*}$	\\
XMM $n_{e}$  2D kT over XMM  $n_{e}$  3D kT  in the $2-5$ keV band 													&	\seciac{lambda_study}		& \figiac{fig:t2dt3d}							&	-2 / +2	\\
CXO dep. $n_{e}$ over XMM dep. $n_{e}$, normalised at $0.15R_{500}$ 											 	&	\seciac{psf_study}				& \figiac{fig:xmm_ne_psftest}			&	-5 / +2	\\
XMM dep. $n_{e}$ computed in the $(2-5)$ keV band over the dep. $n_{e}$  in the $(0.3-2)$ keV band		&	\seciac{psf_comparison}		& \figiac{fig:psf_xmm_energy}			&	-2 / + 3	 \\
\hline
\end{tabular}
}
\end{center}
\footnotesize{Notes: $^*$ The variations are less than $0.5\%$.}
\end{table*}
X-ray observations suffer from background contamination. This background can be divided into two components: instrumental and sky. The first component 
is due to the interaction of the instrument with high energy particles. Despite the filtering processes there is a residual component which contaminates the observations. 
 The sky component is due to the Galactic emission (e.g. \citealt{snowden1995}) and the CXB due to the superposition of unresolved point-sources (e.g. \citealt{giacconi2001}).
When dealing with extended sources which nearly fill the field of view, as for the galaxy clusters, we need dedicated datasets to estimate and subtract the local background.
For \chandra\ and \xmm\ there are two set of datasets. The first one, namely the blank sky dataset, results from the stacking of observations free from diffuse emission where all the point sources have been removed. This dataset represents the average total background (sky plus instrumental). The second set of datasets is formed by specific observations tailored to isolate the instrumental component i.e. block the light coming from the telescope. For \chandra\ this is achieved by moving the instrument away from the focal plane and the resulting observations are the "stowed" datasets. For \xmm\ "closed" datasets are obtained by observing with a configuration of the filter wheel which blocks the light from the mirrors.
In this work we use the latter datasets to remove the instrumental contribution only since this allows us to estimate the sky background component using physically motivated models. That is, using blank sky datasets may result in an overestimate of the local sky background which results in using negative models to account for the over-subtraction\footnote{Another approach is the use of an analytical model for the background (e.g. \citealt{bartalucci2014}).}.

Instrumental background datasets are cleaned following the same criteria used for observations and skycasted to match the same aimpoint and roll-angle. In particular, \chandra\ stowed datasets are selected to match the same background period as the observation (see \citealt{chandra_back} for the background period classification). 
Closed datasets are normalized to the observation in the $[10-12],[12-14]\si{\kilo\electronvolt}$ for MOS$1$,MOS$2$ and PN cameras, respectively. Stowed datasets are normalized in the $[9.5-10.6]\si{\kilo\electronvolt}$ band in order to minimize line contamination at high energy \citep{bartalucci2014}. The same point source defined for the observation is then applied to the instrumental datasets.
The sky component is evaluated differently when dealing with surface brightness or temperature profiles extraction so the subtraction process is described in \seciac{surface_bright} and in \seciac{emission_measure}, respectively.
 
 We also produce datasets accounting for the Out Of Time (OOT) or read out events, which are artefacts related to the presence of very bright sources. We use the method described in \cite{chandra_back} and the \verb?epchain? tool to generate the OOT background datasets for \chandra\ and \xmm, respectively.  These datasets are then processed in the same way as the instrumental background and normalized accordingly. 
Since the instrumental plus particle and OOT datasets are always summed in our analysis from now on we will refer to the two only as the instrumental background. 

It is worth noting that the $V06/V09$ sample is analysed using the \chandra\ blank sky datasets.

\section{\chandra\ weight evaluation}\label{appendix_weight}
Analogously to the \xmm\ \verb?evigweight? tool, we want to evaluate the effective area for each event in function of its precise position (on the detector) and energy also for the \chandra\ dataset. 
We use the CIAO \verb?eff2evt? tool to compute the necessary terms which determine the effective area. To avoid confusion with the tool nomenclature, the outputs for each event are the quantum efficiency (QE), the dead area correction (DACORR) and the "effective area"($\widetilde{EA}$). This latter term is \textit{not} the same as $EA(i,j,e)$ in \eqiac{eq:wght_xmm}. $\widetilde{EA}$ takes into account only effects related to mirrors. The effective area for each event, using this tool, is:
\begin{equation}
EA(i,j,e) = QE(i,j,e) \times DACORR (i,j,e) \times ( \widetilde{EA}(i,j,e),
\end{equation}
where QE and DACORR are unitless quantities and $\widetilde{EA}$ is in cm$^2$.

This tool has been conceived to analyse point sources, so we perform several tests to verify that the tool can be applied also to extended source analysis. The first test is to evaluate the impact of using the energy of the photon, which is affected by instrument degradation problems, to evaluate the effective area. We extract the RMF for each tile of each ACIS CCD in order to carefully map the energy response. For each photon we then determine the most likely energy and compute the $EA(i,j,\tilde{e})$, where $\tilde{e}$ is the corrected energy, and compare to $EA(i,j,e)$ . In the energy range of interest used in this work, $[0.7,10]\si{\kilo\electronvolt}$, we do not find significant differences. An additional test is described in \seciac{par_vs_vig}.

During the observation the \chandra\ telescope is set in a "dithering" motion. The aimpoint insists on different pixels and the resulting effective area changes because of the instrumental effects. For this reason, we can not define the same $EA(i0,j0,e)$ as for \xmm. We compute an average value over time using the \verb?mkarf? tool, which evaluates the effective area at a given Detector position using the appropriate CALDB files and averages over the observation time.

\section{Test summary}\label{appendix_summary}
We report in Tab. \ref{tab:resume_table} the minimum and maximum values of the median of the ratio (i.e. the black solid line in each plot) computed for each test described in \seciac{robustness}. 

\newpage

\begin{figure*}[!htbp]
 \begin{center}
  \scalebox{0.8}{\includegraphics{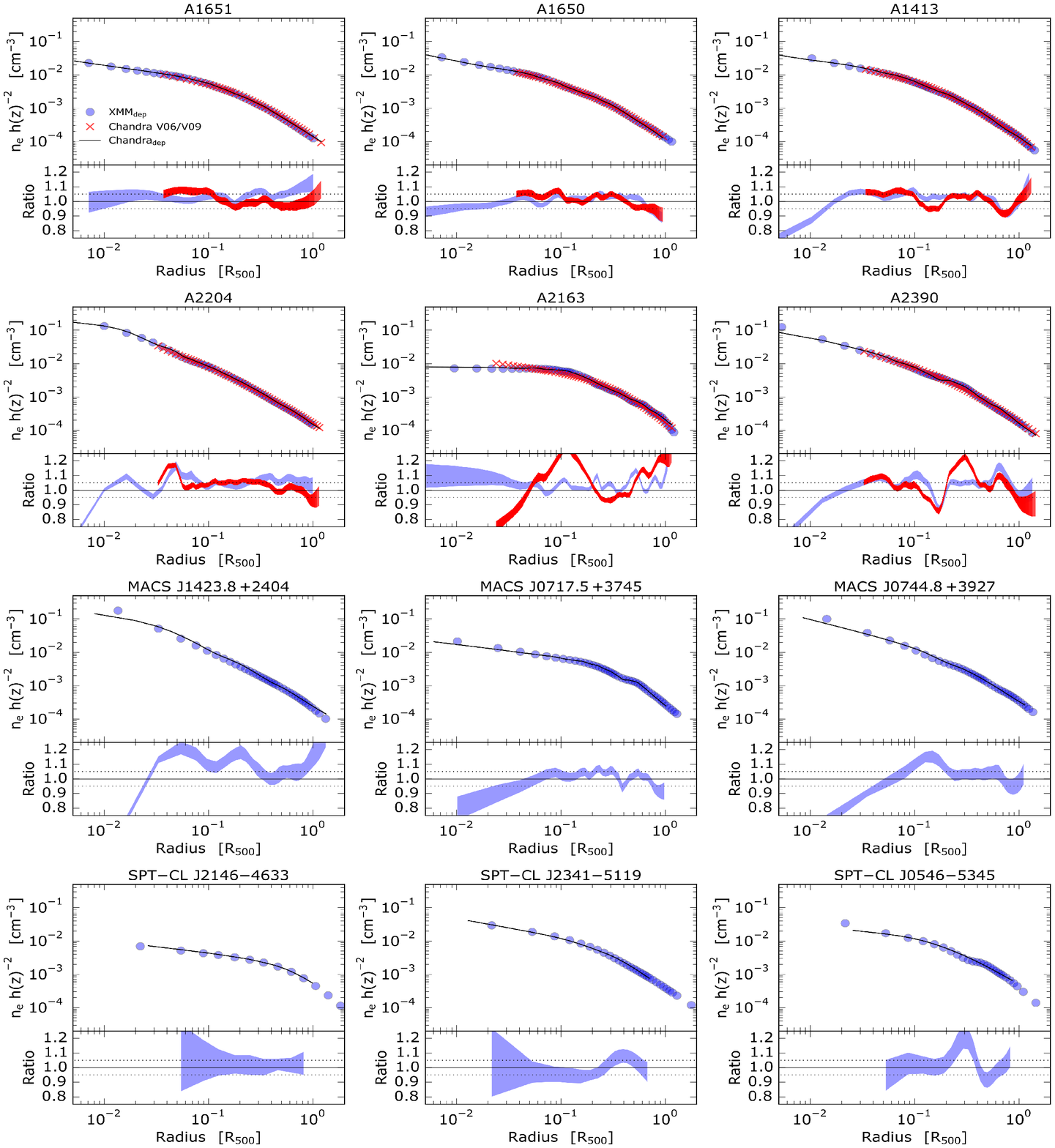}}
 \end{center}
 \caption{\footnotesize {\it For each panel, top:} deprojected scaled density profile obtained using \chandra\ and \xmm\ plotted using black solid line and blue points, respectively. We also show the parametric density profiles published in V06 with red crosses. {\it For each panel, bottom:} ratio between the \chandra\ and the \xmm\ density profiles and between our \chandra\ and V06/V09 samples using blue and red shaded areas, respectively. The solid line and the dotted lines represent the unity and the $\pm 5\%$ levels, respectively.}
 \label{fig:neall}
\end{figure*}

\begin{figure*}[!htbp]
 \begin{center}
   \scalebox{0.8}{\includegraphics{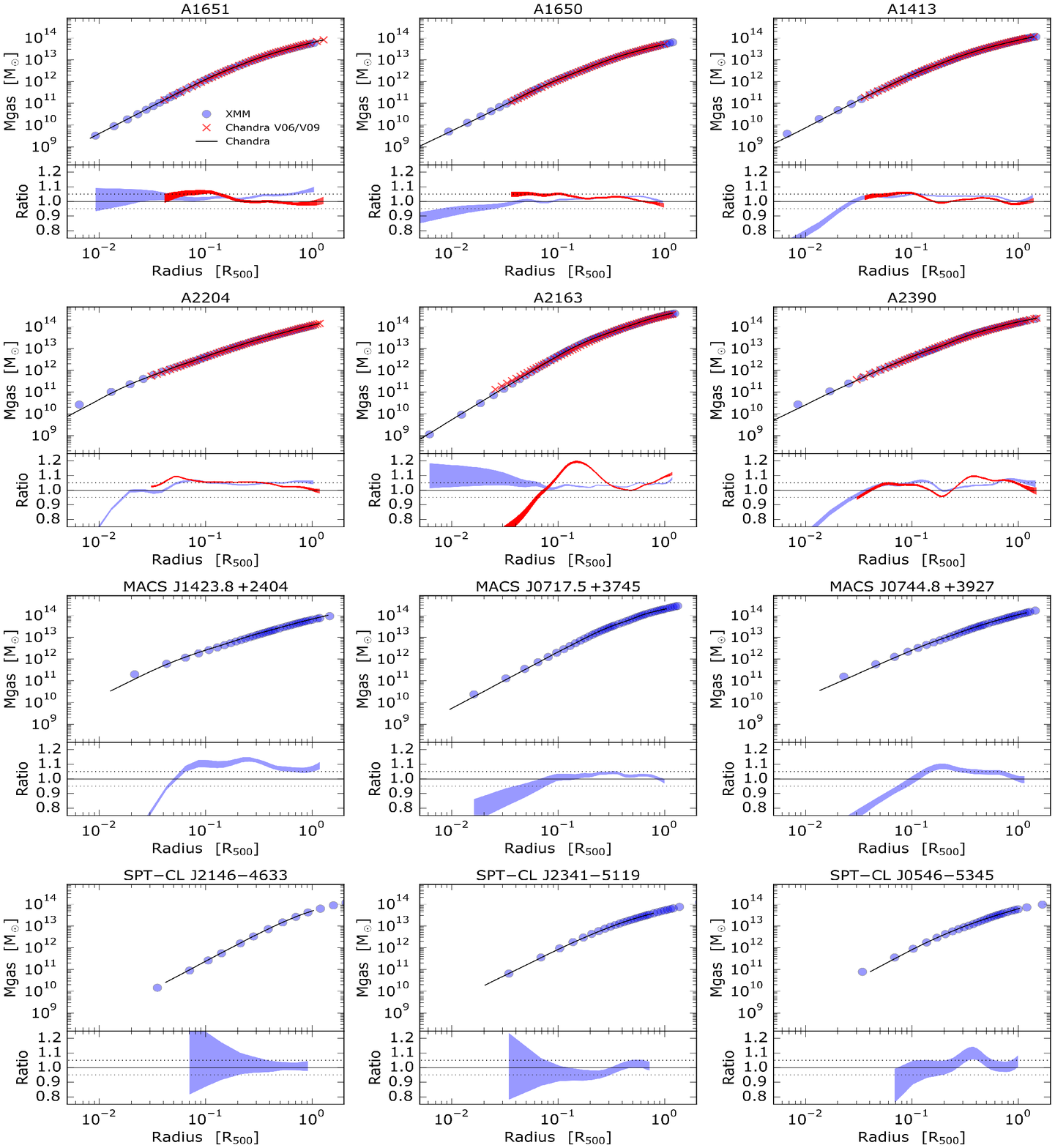} }
 \end{center}
 \caption{\footnotesize Same as \figiac{fig:neall}, except that we show the gas mass profiles.}
 \label{fig:mgasall}
\end{figure*}

\end{document}